\documentclass[twocolumn]{aastex631}

\usepackage{rotating}

\newcommand{\degree}{$^{\circ}$}

\newcommand{\ie}{\textit{i.e.},}
\newcommand{\eg}{\textit{e.g.},}
\newcommand{\survey}{\textit{HST}-Hyperion}
\newcommand{\HST}{\textit{HST}}
\newcommand{\zp}{$z_{\rm phot}$}
\newcommand{\zg}{$z_{\rm grism}$}
\newcommand{\zs}{$z_{\rm spec}$}
\newcommand{\hstmag}{$m_{HST}$}
\newcommand{\grizli}{\textsc{grizli}}
\newcommand{\qfg}{$qf_{\rm g,q}$}
\newcommand{\qfgz}{$qf_{\rm g,z}$}
\newcommand{\qf}{$qf$}
\newcommand{\qfs}{$qf_{\rm s,z}$}
\newcommand{\Halpha}{H$\alpha$}
\newcommand{\Hbeta}{H$\beta$}
\newcommand{\OIIIfive}{\hbox{{\rm [O}\kern 0.1em{\sc iii}{\rm ]$\lambda5007$}}}
\newcommand{\OIII}{\hbox{{\rm [O}\kern 0.1em{\sc iii}{\rm ]}}}
\newcommand{\OII}{\hbox{{\rm [O}\kern 0.1em{\sc ii}{\rm ]}}}
\newcommand{\NII}{\hbox{{\rm [N}\kern 0.1em{\sc ii}{\rm ]}}}

\shorttitle{The \survey\ Survey}
\shortauthors{Forrest et al.}

\graphicspath{{./}{figures/}}

\begin{document}

\title{The \survey\ Survey: Grism Observations of a $z\sim2.5$ Proto-Supercluster}

\correspondingauthor{Ben Forrest}
\email{bforrest@ucdavis.edu}

\author[0000-0001-6003-0541]{Ben Forrest}
    \affiliation{Department of Physics and Astronomy, University of California Davis, One Shields Avenue, Davis, CA, 95616, USA}
\author[0000-0001-9495-7759]{Lu Shen}
    \affiliation{Department of Physics and Astronomy, Texas A\&M University, College Station, TX, 77843 USA}
    \affiliation{George P. and Cynthia Woods Mitchell Institute for Fundamental Physics and Astronomy, Texas A\&M University, College Station, TX, 77843 USA}
\author[0000-0002-1428-7036]{Brian C. Lemaux}
    \affiliation{Gemini Observatory, NSF NOIRLab, 670 N. A’ohoku Place, Hilo, HI, 96720, USA}
    \affiliation{Department of Physics and Astronomy, University of California Davis, One Shields Avenue, Davis, CA, 95616, USA}
\author[0000-0001-7811-9042]{Ekta Shah}
    \affiliation{Department of Physics and Astronomy, University of California Davis, One Shields Avenue, Davis, CA, 95616, USA}

\author[0000-0002-9336-7551]{Olga Cucciati}
    \affiliation{INAF Osservatorio di Astrofisica e Scienza dello Spazio di Bologna, Via Piero Gobetti 93/3, 40129 Bologna, Italy}
\author[0000-0001-8255-6560]{Roy R. Gal}
    \affiliation{Institute for Astronomy, University of Hawai‘i, 2680 Woodlawn Drive, Honolulu, HI 96822, USA}
\author[0009-0003-2158-1246]{Finn Giddings}
    \affiliation{Institute for Astronomy, University of Hawai‘i, 2680 Woodlawn Drive, Honolulu, HI 96822, USA}    
\author[0000-0001-5160-6713]{Emmet Golden-Marx}
    \affiliation{INAF - Osservatorio astronomico di Padova, Vicolo Osservatorio 5, 35122 Padova, Italy}
    \affiliation{Department of Astronomy, Tsinghua University, Beijing 100084, China} 
\author[0000-0003-3424-3230]{Weida Hu}
    \affiliation{Department of Physics and Astronomy, Texas A\&M University, College Station, TX, 77843 USA}
    \affiliation{George P. and Cynthia Woods Mitchell Institute for Fundamental Physics and Astronomy, Texas A\&M University, College Station, TX, 77843 USA}
\author[0000-0001-5749-5452]{Kaila Ronayne}
    \affiliation{Department of Physics and Astronomy, Texas A\&M University, College Station, TX, 77843 USA}
    \affiliation{George P. and Cynthia Woods Mitchell Institute for Fundamental Physics and Astronomy, Texas A\&M University, College Station, TX, 77843 USA}
\author[0000-0001-5796-2807]{Derek Sikorski}
    \affiliation{Institute for Astronomy, University of Hawai‘i, 2680 Woodlawn Drive, Honolulu, HI 96822, USA}
\author[0000-0002-8877-4320]{Priti Staab}
    \affiliation{Department of Physics and Astronomy, University of California Davis, One Shields Avenue, Davis, CA, 95616, USA}

\author[0000-0001-5758-1000]{Ricardo O. Amor\'{i}n}
\affiliation{Instituto de Astrof\'{i}sica de Andaluc\'{i}a (CSIC), Apartado 3004, 18080 Granada, Spain}
\author[0000-0002-8900-0298]{Sandro Bardelli}
    \affiliation{INAF Osservatorio di Astrofisica e Scienza dello Spazio di Bologna, Via Piero Gobetti 93/3, 40129 Bologna, Italy}
\author[0000-0001-7455-8750]{Bianca Garilli}
    \affiliation{INAF-IASF Milano, Via Alfonso Corti 12, 20133 Milano, Italy}
\author[0000-0001-6145-5090]{Nimish Hathi}
    \affiliation{Space Telescope Science Institute, 3700 San Martin Drive, Baltimore, MD 21218, USA}
\author[0000-0001-7523-140X]{Denise Hung}    
    \affiliation{Gemini Observatory, NSF’s NOIRLab, 670 N. A’ohoku Place, Hilo, HI, 96720, USA}    
\author[0000-0003-2119-8151]{Lori Lubin}
    \affiliation{Department of Physics and Astronomy, University of California Davis, One Shields Avenue, Davis, CA, 95616, USA}
\author[0000-0002-3007-0013]{Debora Pelliccia}
    \affiliation{UCO/Lick Observatory, Department of Astronomy \& Astrophysics, UC Santa Cruz, 1156 High Street, Santa Cruz, CA 95064, USA}
\author[0000-0003-0894-1588]{Russell E. Ryan}
    \affiliation{Space Telescope Science Institute, 3700 San Martin Drive, Baltimore, MD 21218, USA}
\author[0000-0002-2318-301X]{Gianni Zamorani}
    \affiliation{INAF Osservatorio di Astrofisica e Scienza dello Spazio di Bologna, Via Piero Gobetti 93/3, 40129 Bologna, Italy}
\author[0000-0002-5845-8132]{Elena Zucca}
    \affiliation{INAF Osservatorio di Astrofisica e Scienza dello Spazio di Bologna, Via Piero Gobetti 93/3, 40129 Bologna, Italy}

\begin{abstract}

We present first results and catalogs from the \survey\ survey.
This survey has collected 50 orbits of WFC3/F160W imaging and WFC3/G141 grism spectroscopy in the most overdense regions of the Hyperion proto-supercluster at $z\sim2.45$, which are analyzed in conjunction with the adjacent 56 orbits of WFC3/F140W imaging and WFC3/G141 grism spectroscopy from the 3D-\HST\ survey.
Sources were identified and spectra extracted using \grizli, which subsequently fit the combined grism data with object-matched photometric data from the COSMOS2020 catalog to obtain a redshift and best-fit spectral model.
Each source was then visually inspected by multiple team members and quality flags were assigned.
A total of 12814 objects with \hstmag~$\leq 25.0$ were inspected, of which 5629 (44\%) have reliable redshifts from the grism data, which are sensitive to emission lines at a level of \mbox{$\sim8.8 \times10^{-18}$~erg s$^{-1}$ cm$^{-2}$} ($1\sigma$).
Comparison to high-quality ground-based spectroscopic redshifts yields a scatter of $\sigma_{\rm NMAD} = 0.0016$.
The resulting catalogs contain 125 confirmed members of the Hyperion structure within $2.40<z<2.53$, with an additional 71 confirmed galaxies in projection within $2.35<z<2.65$.
The redshift, stellar population, and line flux catalogs, as well as all grism spectra, are publicly available.

\end{abstract}

\keywords{Galaxy evolution(594) --- High-redshift galaxy clusters(2007)}

\section{Introduction} \label{Sec:Intro}

While differences between galaxy populations in local clusters and those in the coeval field across a range of properties are quite apparent \citep[\eg][]{Gomez2003,Kauffmann2004,Thomas2005}, the epoch at which these differences first appear is less clear.
To understand the mechanisms and timescales responsible we must investigate the progenitors of galaxy clusters, \ie\ protoclusters.
However, studies of these galaxy protoclusters and galaxy groups, the predominant overdensities in the first few Gyr of cosmic time, are challenging.
These difficulties stem from the wide areas and substantial depths required in multiple photometric passbands to accurately model redshifts for the galaxies comprising these rare objects and supplemental deep spectral observations needed for confirmation.
Finding and characterizing such systems is further complicated by the decreased overdensities of these systems to the coeval field as compared to local clusters \citep[1-2 orders of magnitude dilution;][]{Chiang2013} and their often large spatial extents.

Many individual overdensities have been found, often by following up (potentially biased) signposts such as submillimeter and dusty star-forming galaxies \citep[\eg][]{Daddi2009,Oteo2018}, Lyman-$\alpha$ emitters \citep[\eg][]{Ouchi2005,Lemaux2009,Pavesi2018,Higuchi2019}, active galactic nuclei \citep[\eg][]{Decarli2017,Noirot2018,Mei2023}, and ultra-massive galaxies \citep[\eg][]{McConachie2022, Kakimoto2023, Stawinski2024, deGraaff2024a, McConachie2025}.
Even for systems detected via biased tracers, it is possible to characterize the system in a relatively unbiased manner with significant spectroscopic follow-up.
However, drawing solid conclusions about overdense environments at early times requires large, unbiased samples of such structures, and understanding their effects on galaxy evolution requires highly complete spectroscopic information of member galaxies as well.

While detection of unbiased protocluster samples is difficult, it is possible in some well-sampled extragalactic fields where significant numbers of deep photometric bandpass and spectroscopic observations of the field have been taken.
The Charting Cluster Construction with VUDS \citep{LeFevre2013} and ORELSE \citep{Lubin2009} Survey \citep[C3VO;][]{Lemaux2022} has used such observations to map out overdense systems in the COSMOS, CFHTLS-D1, and ECDFS extragalactic fields, identifying hundreds of candidate systems \citep{Hung2025}.
Additional C3VO spectroscopy from Keck/MOSFIRE and Keck/DEIMOS has resulted in confirmation and characterization of a number of these systems \citep{Cucciati2014, Cucciati2018, Shen2021, Forrest2023, Forrest2024a, Shah2024a, Shah2024b, Staab2024}.

One of these overdensities is the Hyperion proto-supercluster at $z\sim2.45$ \citep{Cucciati2018}.
This system consists of seven overdense peaks, some of which were independently identified previously \citep{Casey2015, Diener2015, Chiang2015, Perna2015, Lee2016, Wang2016}.
Despite extensive spectroscopic observations of the system, large projected areas with numerous photometric member candidates remain.
These regions of low and varied spectroscopic completeness contribute significantly to uncertainties on the effects of the system on the evolution of its member galaxies.

In order to explore this system more comprehensively, we obtained 50 orbits of \HST\ observations (PI: Lemaux, PID 16684), including WFC3/F160W imaging and WFC3/G141 grism spectroscopy in 25 pointings.
The grism data allow for spectroscopic observations of all (sufficiently bright, non-overlapping) galaxies in the field of view.
This work presents the survey dataset, and corresponds to the public data release.
Further works exploring \eg\ the stellar mass function, quiescent fraction, and close kinematic companions will follow.

Herein we detail the reduction of these data (Section~\ref{Sec:Data}) and determination of redshifts and model fits (Section~\ref{Sec:Red}), before comparing the results of redshifts (Section~\ref{Sec:RedComp}) and emission lines (Section~\ref{Sec:EL}) to those from other datasets.
We also provide an updated picture of the Hyperion system (Section~\ref{Sec:Hyp}). 
We assume a $\Lambda$CDM cosmology with $H_0=70$ km/s/Mpc, $\Omega_M=0.30$, and $\Omega_\Lambda=0.70$, and use the AB magnitude system \citep{Oke1983}.

\section{Data} \label{Sec:Data}

\begin{figure*}
    \includegraphics[width=0.8\textwidth, trim=0in 3in 7in 0in]{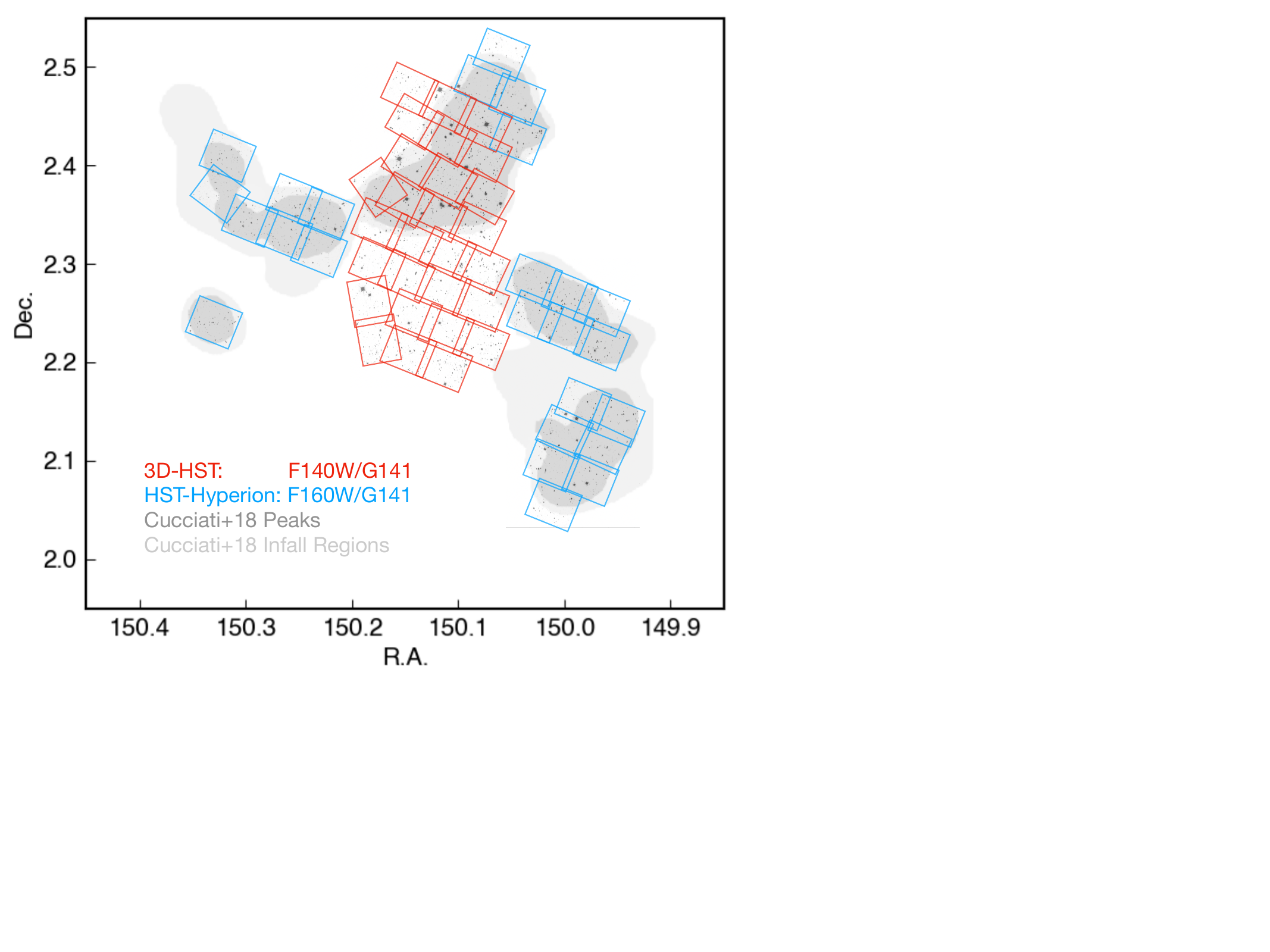}
    \caption{The \HST/WFC3/G141 pointings used in this work. Direct F140W images from the 3D-\HST\ survey are outlined in red, while direct F160W images from the \survey\ survey are outlined in blue. The dark gray shows the 7 peaks of $\geq 5\sigma$ overdensity from \citet{Cucciati2018} (the north and northwest regions each have two peaks overlapping in projection).}
    \label{fig:obsmap}
\end{figure*}

\begin{table*}
	\centering
	\caption{\survey\ Observations. 
    Initial observations of WFC3-PEAK4N5-1 had higher than expected background noise and so were reobserved. All the \HST\ data used in this paper can be found in MAST: 
    \dataset[10.17909/hy1b-pt10]{http://dx.doi.org/10.17909/hy1b-pt10}
    }
	\label{tab:obs}
	\begin{tabular}{lrrl}
		\hline
		Pointing	& R.A. 	& Dec. & Usable G141 Datasets \\
		\hline
		WFC3-PEAK1-1	& 10 00 08.9592 	& +02 25 17.14 	& IENN01021, 01031, 01041, 01051  \\
		WFC3-PEAK1-2	& 10 00 09.4872 	& +02 27 34.81 	& IENN02041, 02051, 52021, 52031  \\
		WFC3-PEAK1-3	& 10 00 17.9520	& +02 28 24.92 	& IENN03021, 03031, 03041, 03051  \\
		WFC3-PEAK1-4	& 10 00 13.4760	& +02 30 08.93 	& IENN04021, 04031, 04041, 04051  \\
		WFC3-PEAK2-1	& 9 59 54.3072 	& +02 09 33.67 	& IENN05021, 05031, 05041, 05051  \\
		WFC3-PEAK2-2	& 9 59 45.9432 	& +02 08 42.06 	& IENN06021, 06031, 06041, 06051  \\
		WFC3-PEAK2-3	& 9 59 58.3800 	& +02 07 43.68		& IENN07021, 07031, 07041, 07051  \\
		WFC3-PEAK2-4	& 9 59 50.1744 	& +02 06 48.66		& IENN08021, 08031, 08041, 08051  \\
		WFC3-PEAK2-5	& 10 00 01.5456 	& +02 05 50.90		& IENN09021, 09031, 09041, 09051  \\
		WFC3-PEAK2-6	& 9 59 53.2584 	& +02 04 59.56 	& IENN10021, 10031, 10041, 10051  \\
		WFC3-PEAK2-7	& 10 00 00.7176	& +02 03 36.26		& IENN61021, 61031, 61041, 61051   \\
		WFC3-PEAK3-1	& 10 00 06.1632 	& +02 16 52.26 	& IENN12021, 12031, 62021, 62031  \\
		WFC3-PEAK3-2	& 10 00 04.9896 	& +02 14 31.38 	& IENN13021, 13031, 63021, 63031  \\
		WFC3-PEAK3-3	& 9 59 57.7728 	& +02 15 59.05 	& IENN14021, 14031, 14041, 14051  \\
		WFC3-PEAK3N7-1	& 9 59 49.3824 	& +02 15 02.60 	& IENN15021, 15031, 65021, 65031  \\
		WFC3-PEAK3N7-2	& 9 59 56.5896 	& +02 13 37.15 	& IENN16021, 16031, 66021, 66031  \\
		WFC3-PEAK3N7-3	& 9 59 48.1632 	& +02 12 42.85 	& IENN17021, 17031, 17041, 17051  \\
		WFC3-PEAK4-1	& 10 01 17.3856	& +02 24 14.02		& IENN18021, 18031, 18041, 18051  \\		
		WFC3-PEAK4-2	& 10 01 18.9984	& +02 22 10.40 	& IENN19041, 19051, 0B021, 0B031  \\
		WFC3-PEAK4-3	& 10 01 12.0048	& +02 20 22.92 	& IENN20021, 20031, 20041, 20051  \\
		WFC3-PEAK4-4	& 10 01 02.2008 	& +02 21 35.05 	& IENN21021, 21031, 21041, 21051  \\
		WFC3-PEAK4-5	& 10 01 03.4560 	& +02 19 35.46 	& IENN22021, 22031, 22041, 22051  \\		
		WFC3-PEAK4N5-1	& 10 00 55.1688	& +02 18 41.57 	& IENN23021, 23031, 23041, 23051, 93021, 93031  \\
		WFC3-PEAK4N5-2	& 10 00 53.7864	& +02 20 43.85 	& IENN24021, 24031, 24041, 24051  \\
		WFC3-PEAK6-1	& 10 01 19.5312	& +02 14 33.55 	& IENN25021, 25031, 25041, 25051  \\
		\hline
	\end{tabular}
\end{table*}

\begin{figure*}
	\includegraphics[width=\textwidth, trim=0in 0in 0in 0in]{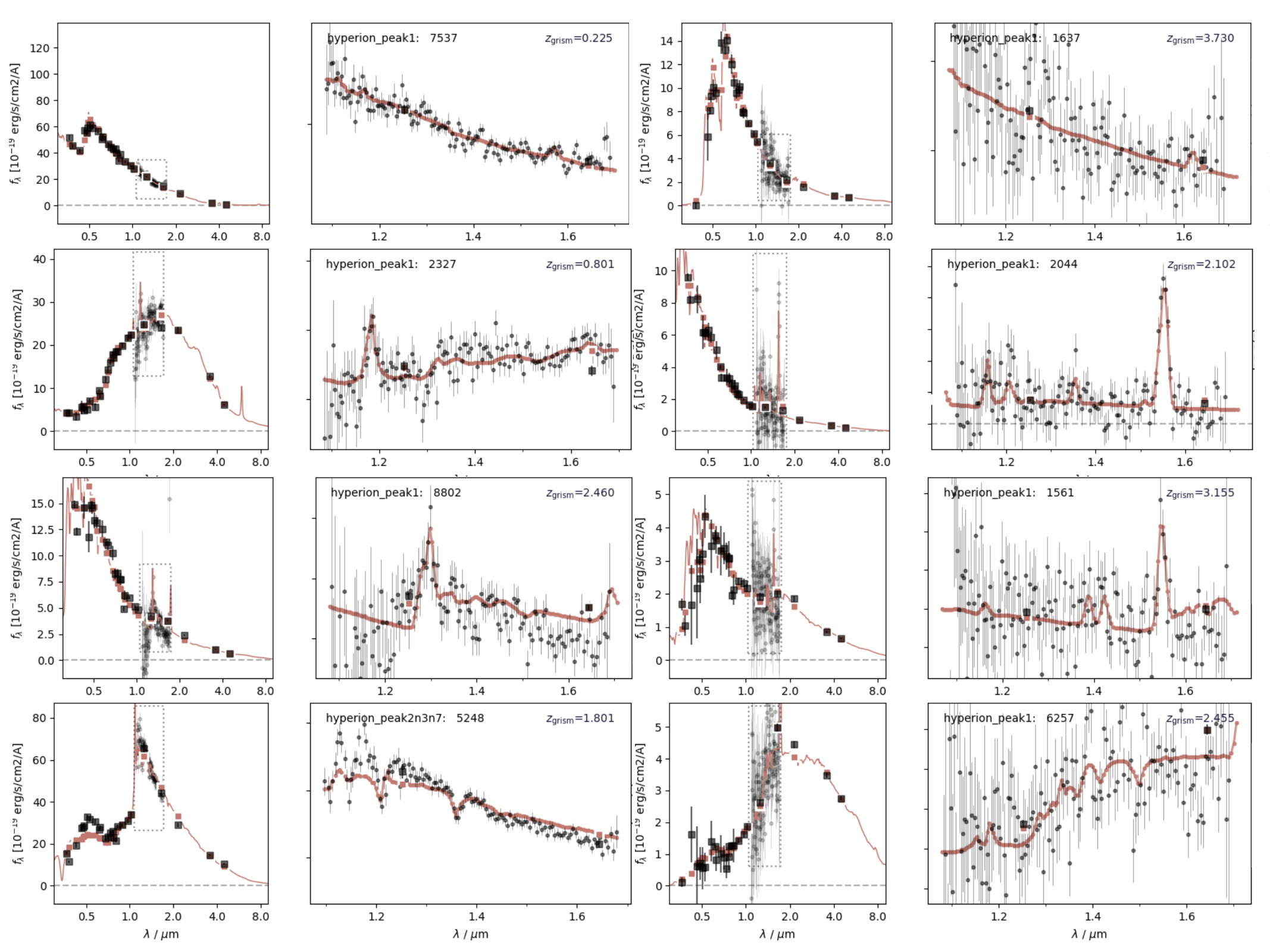}
    \caption{Examples of model fits to observations for eight galaxies in two columns of four, with the left panels highlighting the photometry and the right panels detailing the
    1D extracted grism spectra. Large black points are observed photometry and small black points are observed grism spectra, while the orange line and large points are from the best-fit model. The dotted rectangles on the left panels indicate the grism wavelength regime. This figure shows a variety of spectra, including: continua without obvious emission or absorption features (top row); strong emission lines (\Halpha\ on the left of the 2nd row, \OIII\ on the right of the 2nd row, and \OII\ on the 3rd row); Balmer absorption features (bottom row).}
    \label{fig:exfits}
\end{figure*}

\subsection{\survey\ and Observations}

The \survey\ survey is a Cycle 29 program (PI: Lemaux, PID 16684) consisting of 50 orbits of direct imaging with \HST\ WFC3/F160W and slitless (grism) spectroscopy with \HST\ WFC3/G141 over 25 pointings in the COSMOS field.
The pointing locations and position angles were chosen to complement the 28 pointings of grism observations from 3D-\HST\ \citep[see below;][]{Brammer2012,Momcheva2016} in the COSMOS field, a survey that partially covered the most massive peak of Hyperion, and to target the the other six density peaks of Hyperion identified in \citet{Cucciati2018} as shown in Figure~\ref{fig:obsmap}. 

A nearly identical survey approach to that of the 3D-\HST\ survey was adopted for \survey\ observations. This was done to impose (near) uniformity between the two sets of observations as the two were envisioned to be used in conjunction with each other. Similar to the 3D-\HST\ survey, observations consisted of two orbits per pointing at a single orientation (typically 111\degree\ or 291\degree). Each orbit consisted of four $\sim$553s slitless G141 grism observations bookended by two $\sim$178s F160W direct images. Direct images were placed at the beginning and end of each orbit both to increase the precision of the astrometry and to reduce the effects of the geocoronal HeI line emission in the grism frames. Both sets of images were taken in MULTIACCUM mode, with direct images using an NSAMP=8 and SPARS25 and grism frames using NSAMP=12 and SPARS50. For each pointing, images were dithered by half pixels in the x- and y-axes of the detector to facilitate a $2\times2$ pixel interlacing, to trivialize detector defects, and to improve spatial and spectral resolution. Observations consisted of non-contiguous groupings of pointings, one of which ran abreast of the northern portion of the 3D-\HST\ survey coverage in the COSMOS field. For each group of pointings, pointings were designed to have small overlaps ($\sim$5\%) with adjacent pointings including those from the 3D-\HST\ survey.

The primary difference between the strategy utilized for the \survey\ survey and that of 3D-\HST\ was the adoption of the F160W filter for direct images in the former as compared to F140W for the latter. For 3D-\HST, the choice of F140W was partially due to the fact that F160W coverage in the survey region had already been taken as part of the Cosmic Assembly Near-infrared Deep Extragalactic Legacy Survey (CANDELS, \citealt{Grogin2011, Koekemoer2011}). No such imaging existed in the regions targeted as part of the \survey\ survey and the F160W band was preferred because it probes a rest-frame wavelength range completely redward of the 4000\AA\ break at the redshift of Hyperion. 

The 3D-\HST\ survey was designed with a single roll angle per two-orbit pointing. This is a seemingly poor approach to adopt for the \survey\ observations as we are purposely pointing at the densest regions of Hyperion where source confusion (i.e., grism trace overlap) may be exacerbated relative to the rarefied field. However, extensive testing of the 3D-\HST\ data showed no decrease in the redshift recovery rate of sources located in pointings that overlap with the densest regions of the most massive peak of Hyperion as compared to sources in pointings that targeted regions well outside of Hyperion. The lack of such a decrease is likely a result of the relative sparsity of protocluster environments, which are more than an order of magnitude less rich than their lower redshift descendents (e.g., \citealt{Chiang2013}), and the efficacy of techniques used to decompose overlapping spectra. As such, in deference to uniformity of approach between the two surveys, we adopted the same single orientation approach for the \survey\ survey. 

The \survey\ data were observed beginning in December 2021 and observed over the course of several years, with the final observations taken in December 2024.
Seven pointings in total required re-observations due to pointing drift issues with the initial observations that came as a result of fine guidance sensor issues. 
In all cases, these re-observations were taken, in some instances multiple times due to repeated fine guidance sensor drifts, and all pointings were ultimately observed successfully. 
Frames that were affected by fine guidance sensor drifts were removed from all subsequent analysis. 
The final observations used in this work are listed in Table~\ref{tab:obs}.

\subsection{Previous Observations in the Literature}

In order to optimally characterize the Hyperion structure and its effects on member galaxies, we wish to utilize as much photometry and spectroscopy of protocluster galaxies, as well as foreground and background sources, as possible.

\citet{Cucciati2018} first identified and characterized the Hyperion proto-supercluster as a whole using data from field spectroscopic surveys including the VIMOS Ultra-Deep Survey \citep[VUDS][]{LeFevre2013} and the zCOSMOS survey \citep{Lilly2007}, as well as photometric data from the COSMOS2015 catalog \citep{Laigle2016}.
This work estimated the total mass of the Hyperion system as $\sim4.8\times 10^{15}$~M$_\odot$ extended over a volume of $\sim5.4\times10^5$~Mpc$^{3}$.
Constrained simulations suggest that this system will collapse into a large filamentary supercluster with four component clusters \citep{Ata2022}.

In this work we consider further field spectroscopy from the DEIMOS10k survey \citep{Hasinger2018} and fold in information from the COSMOS2020 CLASSIC photometric catalog \citep{Weaver2022}.
Additionally, the C3VO survey performed targeted spectroscopic follow-up of Hyperion with Keck/MOSFIRE (see Table 1 of \citealt{Forrest2024a}).
Seven masks were observed in $\sim0.7$\arcsec\ seeing, with a total of 58 new confirmed member redshifts.

\subsection{Data Reduction with \grizli} \label{Sec:DRgrizli}

We reduced the \survey\ data and also re-reduced the 3D-\HST\ data to ensure uniformity.
These data are processed simultaneously, and in what follows, we consider these data together as \survey\ and do not differentiate between the pointings.
However, we do compare redshifts and emission line fluxes and equivalent widths derived for objects in the 3D-\HST\ pointings as originally computed by the 3D-\HST\ team and as computed by the \survey\ team after re-reducing the data.
As such, we recommend that citations of this work also include references to 3D-HST \citep{Brammer2012, Momcheva2016}.

We processed the \HST\ grism data using \grizli\ version 1.9.5 (\citealt{Brammer2021a}; see also \citealt{Momcheva2016} and \citealt{Simons2021} for details). 
\grizli\ performed processing of \HST/WFC3 imaging and slitless spectroscopic data sets, including data retrieval, pre-processing of the raw observations for cosmic rays, flat field corrections, sky subtraction, astrometric corrections, and alignment. 
The F160W images from \survey\ and F140W images from 3D-\HST\ were combined to build a full F140W+F160W mosaic. 
\grizli\ utilizes \textsc{sep} \citep{Barbary2016} to provide source detection, extraction, and segmentation. 
We adopted the default detection parameters with a threshold of 2 times the RMS and a minimum area of 9 pixels. 
To eliminate contamination from nearby sources in each grism spectrum, we generated a contamination model for each \HST\ pointing by forward modeling the \HST\ F140W+F160W full-field mosaic.
Firstly, we created models with constant flux density (flat models) set by either their F140W or F160W magnitudes (hereafter \hstmag), a choice that depended on whether the source fell in the \survey\ ($F160W$) or the 3D-\HST\ ($F140W$) footprint, for all sources with $m_{HST}<26$.
 After subtracting the initial flat models,  we then constructed third-order polynomial models for all sources with \hstmag~$<25$.
We extracted 1D and 2D spectra for all sources with magnitudes in the range \mbox{$18<$~\hstmag~$<26$}.

Grism redshifts were determined simultaneously by fitting the contamination-subtracted grism spectra and available multi-wavelength photometry. 
For the multi-wavelength photometry, we adopted the COSMOS2020 CLASSIC catalog \citep{Weaver2022}. We cross-matched sources detected in our full F140W+F160W mosaic to COSMOS2020 using a 0.5\arcsec\ search radius. 
We removed sources without a match, amounting to $\sim$24\% of \grizli\ sources, the vast majority of which are fainter than \hstmag~$>24$.
Of the 29 sources with \hstmag~$>24$ these are a combination of blended objects, targets with complex morphologies, and sources in masked regions of the COSMOS2020 catalog.
The F140W/F160W photometric data from 3D-\HST/\survey\ were not included in the fits as it was not extracted in the same way as the COSMOS2020 photometry. Note that the wavelength coverage of F140W/F160W is mostly overlapped with UltraVISTA $H$-band (3$\sigma$ depth over 2\arcsec\ aperture $\geq24.9$), and thus excluding these \HST\ data does not result in a significant loss of information. 
The spectroscopic data was scaled to match the photometric data with a normalization factor based on the best-fit model. 
We adopted a set of templates from the Flexible Stellar Populations Synthesis models \citep[FSPS;][]{Conroy2010} and four combinations of nebular emission lines with fixed line ratios including 1) \Halpha+NII+SII+SIII+He+Pa$\beta$, 2) OIII+H$\beta$+H$\gamma$+H$\delta$, 3) OII+Ne, and 4) a series of UV emission lines. 
A 2D spectral model for each redshift was computed by projecting templates to the pixel grid of the 2D grism spectra using the morphology from \HST\ F140W+F160W imaging.  
We adopted a redshift range of $0 < z < 6$ with a logarithm step in redshift of 0.004 and 0.0005 for the second pass. 
\grizli\ determines a redshift by minimizing the $\chi^2$ between the 2D spectral models and extracted 2D spectra, as well as between the photometric models and photometric data (see more details in \citealp{Momcheva2016}). 

Subsequent to obtaining a redshift, the data were refit with emission line ratios allowed to vary to better match the grism data.
Individual emission-line fluxes were measured using the best-fit redshift and accounting for the continuum flux using the best-fit stellar population model. 
Example fits are shown in Figure~\ref{fig:exfits}.

\section{Redshift Determination}\label{Sec:Red}

\begin{figure*}
	\includegraphics[width=\textwidth, trim=0in 0in 3in 0in]{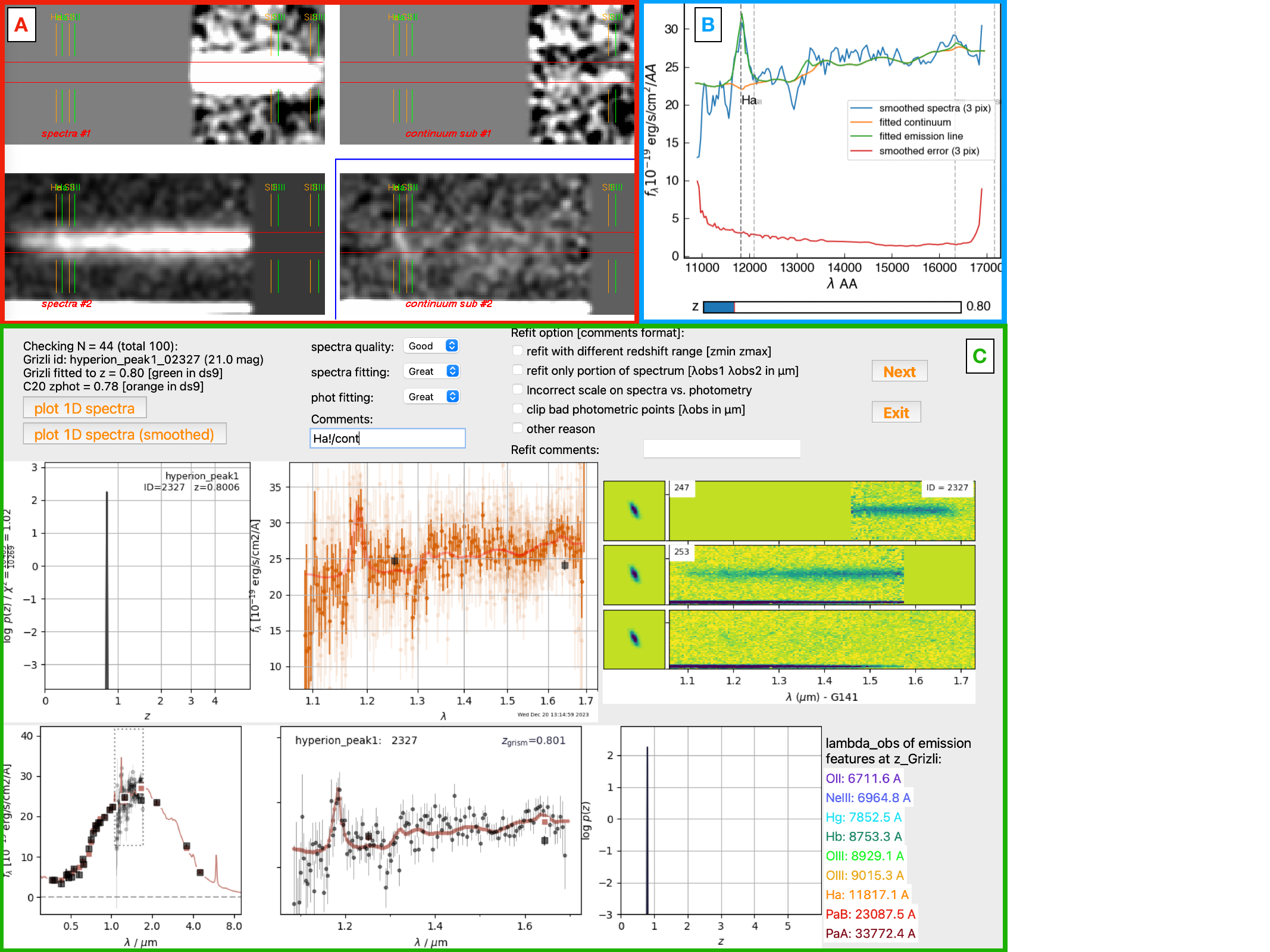}
    \caption{The GUI used for visual inspection of redshift fits and assigning quality flags. \textbf{Panel A:} An interactive ds9 window showing all grism pointings of a specific target (left) and the model continuum subtracted residuals (right). \textbf{Panel B:} An interactive view of the grism spectrum smoothed to the resolution of the G141 instrument. A slider allows the user to see how various emission lines change position with redshift. \textbf{Panel C:} The main GUI window shows output plots from \grizli\ including the model fit to the 1D grism spectrum and photometry, the resultant redshift probability distribution, and the 2D grism spectrum.}
    \label{fig:gui}
\end{figure*}

\subsection{Individual Visual Classification of Initial \grizli\ Fits}

All objects with \hstmag~$\leq 25.0$ were visually inspected by at least two people on a team of twelve using a custom graphical user interface (GUI) developed by one of the co-authors (L.S.; Figure~\ref{fig:gui}).
The \HST\ imaging, original and continuum-subtracted 2D G141 spectra, optimally extracted and flux corrected 1D spectrum, model fit to the 1D spectrum, model fit to the photometry, and the COSMOS2020 \zp\ (if applicable) were all considered and the object was then given three quality flags by each classifier: 1) a spectral quality flag (SQF),  2) a spectral fit flag (SFF), and 3) a photometric fit flag (PFF).
An option to perform a joint refit of the photometry and grism spectrum over either a specific redshift range or using only a portion of the wavelength coverage of the grism was allowed.
Finally, comments on observed spectral features, possible contamination, and labeling of particularly uncertain or borderline cases were also encouraged.

{Broad guidelines were provided for classification, with each flag assigned as Great, Good, Unclear, or Bad.
For the SQF, Good and Great flags required detection of continuum and/or emission lines with little to no contamination. 
Objects with SQF=Unclear either had no clear detection of the source in the G141 observations or a low signal detection combined with weak contamination or other reduction artifacts.
SQF=Bad indicated significant contamination in the spectral extraction window that minimize the usability of the grism spectrum.

The SFF indicates how well the model spectrum (and redshift) assigned by \grizli\ matched the observed grism spectrum, and confidence in the associated redshift.
As a result, SFF=Great required an unambiguous fit, typically well-detected continuum with emission lines and/or a clear continuum break.
SFF=Good indicated a redshift constrained by emission line fits but an ambiguous or poor match to the observed continuum, or cases where a featureless continuum was clearly detected and no emission lines were expected given the fit redshift and/or the shape of the SED.
SFF=Unclear was assigned to objects with noisy grism observations of the continuum or objects where an emission line was expected given the shape of the SED, but not observed, provided that the model was consistent with these observations.
SFF=Bad indicated fits which were obviously incorrect due to poor fitting to observed emission lines or continuum.

The PFF speaks to the goodness-of-fit of the assigned model to the observed photometry, and is Great if all the observed photometry is consistent with the model with the expectations of Gaussian scatter and known zero-point issues.
Objects with PFF=Good had some photometric points which were more significantly offset from the model, but were for the most part consistent.
PFF=Unclear indicated that a conclusion was not apparent as either the photometry had large uncertainties, or that another model/redshift was suspected to provide a better fit.
PFF=Bad indicated a model which clearly did not fit the photometric observations.

All sources were inspected blindly, that is classifiers did not know whether a source was originally observed by \survey\ or 3D-\HST\, and if the latter, any previous redshift information was not used.

\begin{figure*}
	\includegraphics[width=\textwidth, trim=0in 2.5in 0in 0in]{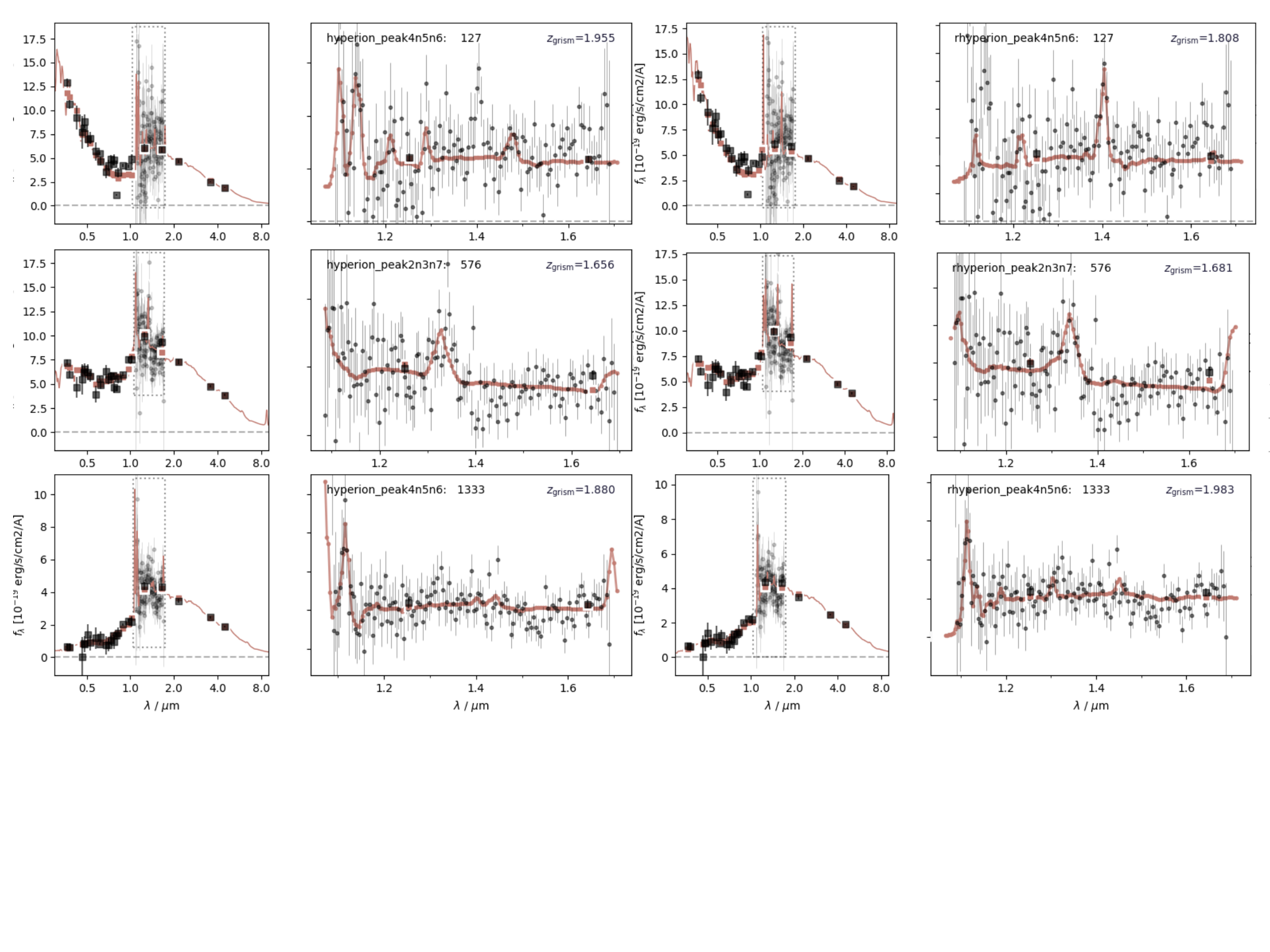}
    \caption{Examples of galaxies that were flagged for a refit following visual inspection of the initial \grizli\ redshift fit. The initial fit is on the left and the result of the refit is on the right. Symbols are the same as in Figure~\ref{fig:exfits}. From top to bottom, the refits were requested to address: 1) the apparent emission line at $\lambda/\mu m\sim1.4$ which was not initially fit, 2) a small redshift adjustment to better fit the photometry and the emission line at $\lambda/\mu m\sim1.33$, and 3) to fit the emission line at $\lambda/\mu m\sim1.12$ as \OII\ instead of \hbox{{\rm [Ne}\kern 0.1em{\sc iii}{\rm ]}}. In these cases, \grizli\ was only allowed to fit over a narrow redshift range specified by the classifier. All refits were subsequently inspected to ensure reasonableness.}
    \label{fig:refits}
\end{figure*}

\begin{figure*}
	\includegraphics[width=\textwidth, trim=0in 0in 0in 0in]{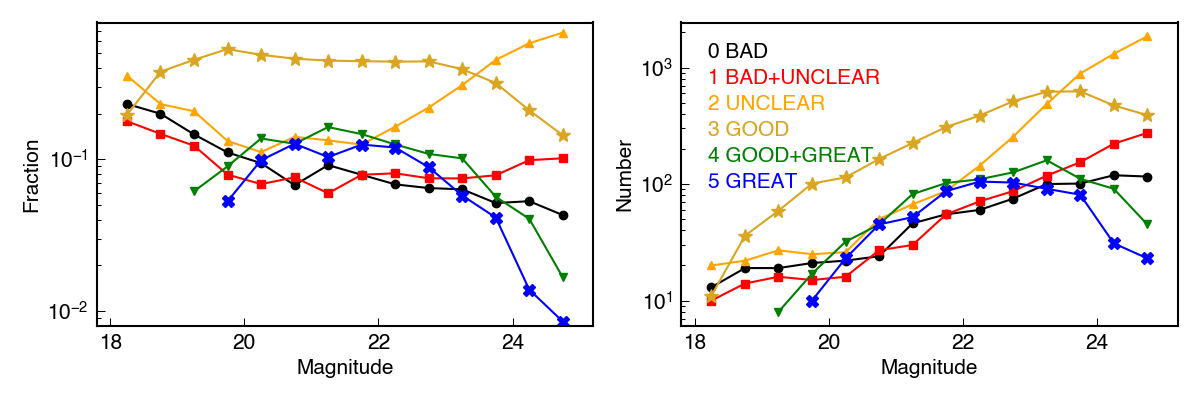}
    \caption{Distribution of redshift quality flags (\qfgz) as a function of WFC3 magnitude. Each colored line denotes a separate redshift quality flag: black circles (0, bad); red squares (1, bad+unclear); orange upward triangles (2, unclear); gold stars (3, good); green downward triangles (4, good+great); blue exes (5, great). The left panel shows the fraction of galaxies in a magnitude bin which were assigned a particular flag, while the right panel shows raw numbers. The number of galaxies with high-quality redshifts (\qfgz~$\geq3$) peaks at $m\sim23.5$.}
    \label{fig:magflag}
\end{figure*}

\subsection{Reconciliation, Refit, and Reinspection}

When there were discrepancies in the binary use flags between classifiers, those classifiers subsequently met to discuss these objects with the goal of reaching a consensus on the binary use flag via changes to the PFF and SFF.
If no changes were agreed upon, the classifiers would keep their original designations.
As a result, each classifier produced a reconciled list of flags which were used in the final flagging scheme.
The total time in person-hours for initial inspection and reconciliation for the entire program is estimated at 850 hours.

When a request for a refit of an object's data at a particular redshift was made by a majority of classifiers, such a refit was performed and these objects were then inspected by an expert classifier (B.F., B.C.L., E.S., or L.S.) to verify if the new redshift was a good fit using the same flagging system.
Similarly, all cases labelled `borderline' after reconciliation were reinspected by an expert classifier.
For such objects, only the flags from the expert classifier were used in the final flagging scheme.
Examples of object fits before and after refitting are shown in Figure~\ref{fig:refits}.

\subsection{Final Quality Flags}

In the final catalog, each object was assigned a grism spectral quality flag (\qfg) and a grism redshift quality flag (\qfgz) indicating the usability of the data and reliability of the assigned redshift, respectively.
Each classifier's SFF and PFF were considered and the worst value was assigned as their individual \qfgz\ ranking.
Meanwhile, each classifier's SQF was assigned as their individual \qfg\ ranking. 
These individual rankings were then combined as follows to obtain the final flags.

In the case of objects with refits requested by a majority of classifiers, and in borderline cases, the only ranking used was that of the expert classifier.
If this ranking was Great (\ie\ both the SFF and the PFF were Great), then \qf=5.
If this ranking was Good (\ie\ both the SFF and the PFF were Good, or one was Great and one was Good), then \qf=3.
If this ranking was Unclear (\ie\ either SFF or PFF was Unclear, and the other was Unclear, Good, or Great), then \qf=2.
If this ranking was Bad (\ie\ either the SFF or PFF were Bad), then \qf=0.

Most objects had two classifiers, and their rankings were combined to calculate the \qf.
If both rankings were Great, \qf=5.
If one ranking was Good and one was Great, \qf=4.
If both rankings were Good, \qf=3.
If one ranking was Unclear, and the other was Unclear, Good, or Great, \qf=2.
If one ranking was Unclear and one was Bad, \qf=1.
If both rankings were Bad, \qf=0.

When objects had more than two classifiers, the modal ranking was used as in the case of one classifier.
If the rankings were bimodal, the two ranking modes were used as in the case of two classifiers.
Otherwise, the lowest ranking from all classifiers was used as in the case of one classifier.
The distribution of final \qfgz\ flags as a function of magnitude is shown in Figure~\ref{fig:magflag}, and the distribution in magnitude-redshift space is shown in Figure~\ref{fig:magflagz}.
While there are clear redshift and magnitude dependencies, the percentage of objects with \qfgz$=0-5$ is 6.19\%, 8.67\%, 41.21\%, 31.49\%, 7.34\%, and 5.10\%, respectively.

\begin{sidewaysfigure*}
    \includegraphics[width=0.98\textwidth, trim=0in 0in 0in 6in]{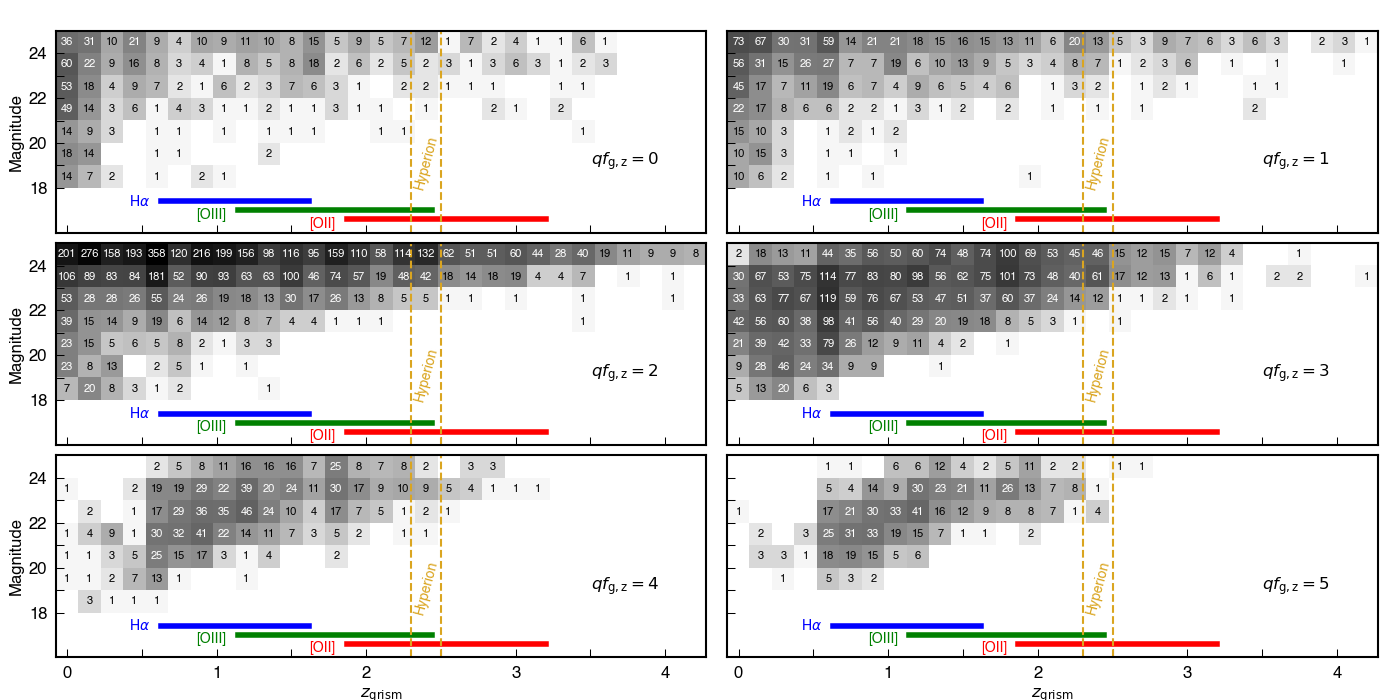}
    \caption{The number of galaxies with an assigned \qfgz\ in bins of grism redshift and WFC3 magnitude. Each bin is shaded by the logarithm of the number of contained galaxies. The redshifts at which the prominent emission lines fall into the G141 wavelength range are labeled, as is the Hyperion redshift. }
    \label{fig:magflagz}
\end{sidewaysfigure*}

\section{Redshift Catalog and Comparison of Redshifts}\label{Sec:RedComp}

\subsection{Object Matching}

As mentioned in Section~\ref{Sec:DRgrizli}, each grism source is matched to a COSMOS2020 CLASSIC photometric entry by finding the closest entry within a projected 0.5\arcsec\ radius.
In this manner there are 12760 out of 12814 (99.6\%)
grism sources with $m_{HST}\leq 25.0$ matched to 12641 COSMOS2020 CLASSIC entries.
While 54 grism sources do not have a corresponding photometric match, there are also 231 grism sources matched non-uniquely to 114 COSMOS2020 sources.
We confirm that there are no other nearby COSMOS2020 sources for these objects, and therefore reason that these grism sources are either: 1) not detected in COSMOS2020, 2) blended together into a single object in COSMOS2020, or 3) not real sources (\ie\ spurious noise detections).

The corresponding $3\sigma$ depths of the COSMOS2020 catalog are $m_J\sim25.9$ and $m_H\sim25.5$, which suggests that sources with \hstmag$\leq 25.0$ should be detected without issue. 
Visual comparison of the \HST\ direct images and ground-based NIR images also confirms that many of the sources in question are indeed blended together.
Of the 231 grism sources, 57 have a reliable redshift (\qfgz~$\geq 3$), of which 32 comprise 16 pairs.
For these 16 pairs of grism sources with reliable redshifts, 14 of the pairs have $\Delta$\zg$<0.03$.
Therefore, we conclude that these grism sources with reliable redshifts are mostly blended pairs.
In this work we do not make any attempt to deblend the photometry associated with these objects and instead mark them in the final redshift catalog with a blending flag.

We use the COSMOS2020 matches found above to match each grism source to the C3VO internal spectroscopic catalog constructed in the COSMOS field detailed in \citet{Forrest2024a}.
Briefly, this catalog matches spectra from C3VO, VUDS, zCOSMOS, and DEIMOS10k to the COSMOS2020 catalog and takes the highest quality redshift for objects with multiple spectra as determined from various survey quality flags and a survey hierarchy.
The final flags are similar to the VUDS flagging system.

\begin{table}
	\centering
	\caption{Columns in the \survey\ redshift catalog.}
	\label{tab:zcat}
	\begin{tabular}{ll}
		\hline
		Column	& Description \\
		\hline
            RA      &   Right Ascension of the grism source (J2000)\\
            DEC     &   Declination of the grism source (J2000)\\
            mag     &   \hstmag\ of the grism source\\
            ID\_C20 &   COSMOS2020 CLASSIC match ID\\
            zp\_lp  &   LePhare \zp\ of COSMOS2020 match (lp\_zBEST)\\  
            zp\_ez  &   EAZY \zp\ of COSMOS2020 match (ez\_z500)\\
            zs      &   Spectroscopic redshift\\  
            qf\_s   &   Spectroscopic redshift quality flag\\
            ID\_g   &   Target ID in \survey \\               
            zg      &   Grism redshift from \survey \\
            zg\_16   &   $16^{\rm th}$ percentile of the final $p(z)$ distribution\\
            zg\_84   &   $84^{\rm th}$ percentile of the final $p(z)$ distribution\\
            qf\_gq  &   Grism data quality flag\\
            qf\_gz  &   Grism redshift quality flag \\
            spec\_com &   Grism classifier comments\\
	\end{tabular}
\end{table}

\begin{figure*}
	\includegraphics[width=\textwidth, trim=0in 0in 0in 0in]{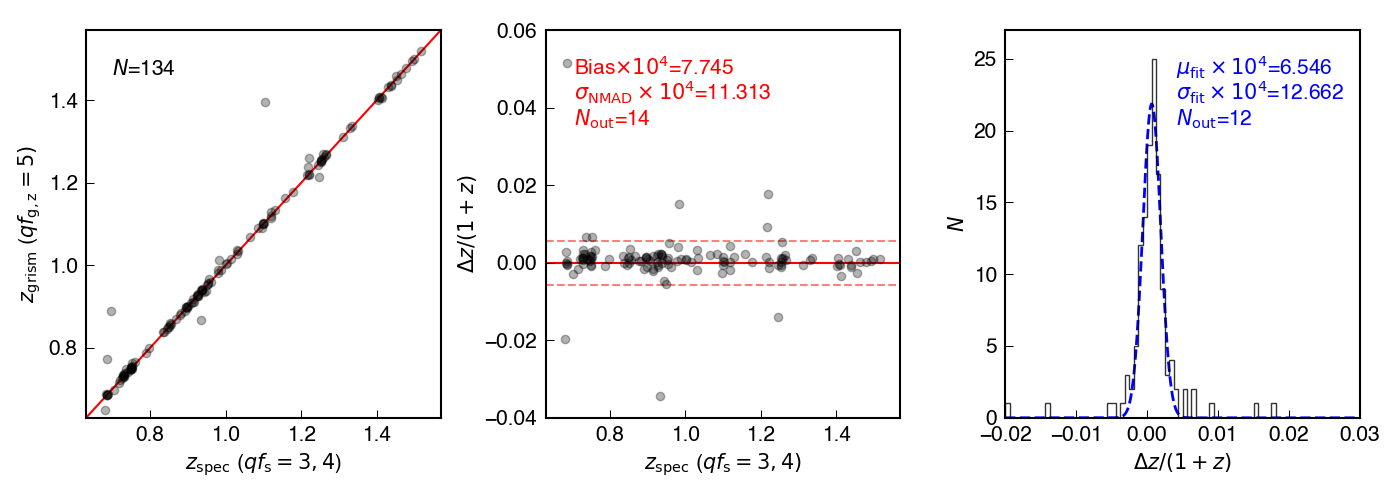}
    \caption{Comparison of high-quality spectroscopic and grism redshifts, demonstrating the excellent agreement between the two. The solid red lines show 1:1 agreement, while the dashed red lines in the center panel represent $\pm5 \sigma_{\rm NMAD}$, outside of which galaxies are considered outliers. On the right panel is a Gaussian fit to $\Delta z / (1+z)$, which results in similar quantification.}
    \label{fig:wavecal}
\end{figure*}

\subsection{Wavelength Calibration Check}\label{sec:wave}

Objects with the highest quality grism redshifts and highest quality ground-based spectroscopic redshifts from the literature were compared to quantify the precision and accuracy of the grism wavelength solution calibration.
For this comparison we only consider objects with ground-based spectroscopic redshift quality flag (\qfs) equal to 3 or 4, corresponding to accuracy rates of $\sim95$\% and $\sim100$\%, respectively, as well as grism redshift quality flag (\qfgz) equal to 5, indicating that multiple classifiers labeled the redshift as `Great'.
Furthermore, we only analyze objects in the redshift range $0.68\leq z \leq 1.52$ for this process, resulting in 134 objects.
Such objects have the \Halpha\ emission line in the grism wavelength window as well as \OII\ emission detected in spectra from either the VIMOS instrument \citep{LeFevre2003} used in the zCOSMOS and VUDS surveys or the DEIMOS instrument \citep{Faber2003} used in the DEIMOS10k survey and some C3VO follow-up.
The comparison of the redshifts determined from these lines removes the uncertainty introduced when comparing redshifts from Ly$\alpha$ emission, which can be significantly offset from the systematic redshift \citep[\eg][]{Steidel2018, Cassata2020}.

As shown in Figure~\ref{fig:wavecal}, there is excellent agreement between the spectroscopic and grism redshifts for this sample of galaxies.
The median offset of \mbox{$\Delta z/(1+z)\sim7\times 10^{-4}$} corresponds to an observed wavelength difference of 9\AA\ for \Halpha\ at $z=1$ or \OII\ at $z=2.45$, the redshift of Hyperion.
The scatter about the median is $\sigma_{\rm NMAD}=0.0011$, slightly lower than that derived from previous grism analyses \citep{Morris2015, Momcheva2016}.
Given that the dispersion of the G141 grism is 46.5\AA\ per pixel, we do not apply any wavelength offset correction to the grism redshifts.

\subsection{Redshift catalog}

The final \survey\ redshift catalog contains 12814
objects with $m_{HST}\leq 25.0$ and associated photometric, spectroscopic, and grism redshift properties, all of which are listed in Table~\ref{tab:zcat}.

The \survey\ ID, right ascension, declination, \hstmag, grism redshift, grism data quality flag, and grism redshift quality flag from \survey\ are included, as are classifier comments on observed emission lines and possible contamination.
Also provided are the COSMOS2020 CLASSIC photometric match IDs and associated \zp\ values from LePhare and EAZY.
Finally, for objects with spectroscopic observations, the 
\zs\ and spectroscopic quality flag are given.

Additionally, we remodel the physical properties of galaxies in the redshift catalog using LePhare \citep{Arnouts1999, Ilbert2006}.
The same parameters and inputs are used as in \citet{Ilbert2009}, COSMOS2020 \citep{Weaver2022}, and \citet{Forrest2024a}, with the redshift fixed to the highest quality redshift, either \zg\ or \zs, based on the confidence level of the associated quality flags.
We include the best-fit stellar mass, star-formation rate, and stellar age for each galaxy in a separate, line-matched catalog.

\begin{figure}
    \includegraphics[width=0.45\textwidth, trim=0in 0in 0in 0in]{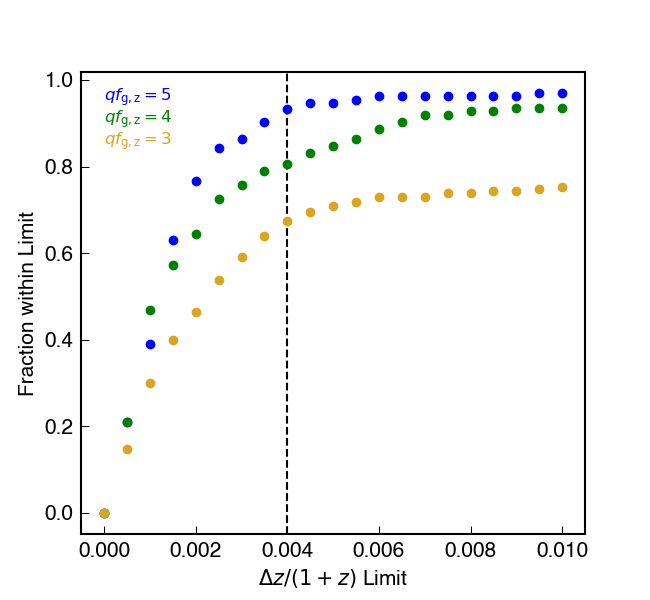}
    \caption{The cumulative distribution functions of $\Delta z / (1+z)$ for sources with high quality spectroscopic redshfits (\qfs~=3,4) and high quality grism redshifts (\qfgz~=3,4,5). These curves start to asymptote around $\Delta z / (1+z)=0.004$, which we use as our cut for determining grism redshift reliability.}
    \label{fig:dzlim}
\end{figure}

\begin{figure*}
	\includegraphics[width=\textwidth, trim=0in 0in 0in 0in]{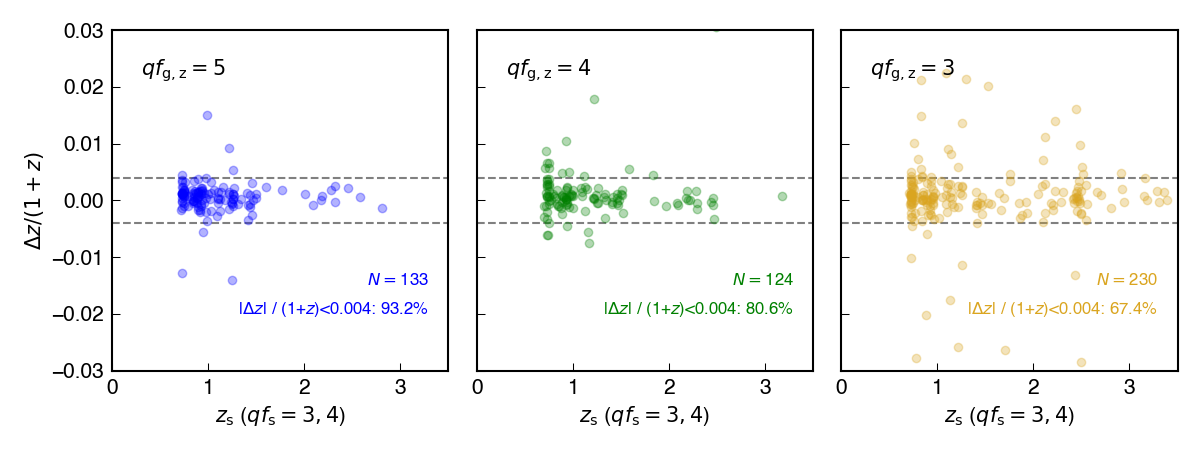}
    \caption{A comparison of redshifts for sources with high quality spectroscopic redshfits (\qfs~=3,4) and high quality grism redshifts (\qfgz~=~3,4,5 from right to left). The percentage of sources with the agreement limit $\Delta z/(1+z)<0.004$ is given. }
    \label{fig:reliability}
\end{figure*}

\subsection{Comparison to Photometric and Spectroscopic Redshifts}\label{Sec:zg_comp}

In Section~\ref{sec:wave} we compared grism and spectroscopic redshifts for a subset of very high quality objects with emission lines probing systematic redshifts (\ie\ excluding Ly$\alpha$).
Here we compare larger samples of grism redshifts with varying quality flags to obtain estimates of grism redshift reliability.
 
In particular, we compare objects with \qfgz~$=3,4,5$, \qfs~$=3,4$, and redshifts \zs~$>0.7$ to remove objects with featureless continua in the grism.
This results in a sample of 133 galaxies with \qfgz=5, 124 galaxies with \qfgz=4, and 230 galaxies with \qfgz=3.
We consider objects with $|z_{\rm grism}-z_{\rm spec}| / (1+z)<0.004$ to be accurate redshifts.
This is motivated by finding the value where the cumulative distribution function of matches begins to asymptote, as shown in Figure~\ref{fig:dzlim}.
This also corresponds to $\Delta \lambda_{\textrm [OII], obs}\sim50{\textrm \AA}$ at the redshift of Hyperion, roughly the dispersion per pixel of the G141 grism.
The fraction of galaxies within this limit is 93.2\% for \qfgz=5, \textbf{80.6\%} for \qfgz=4, and 67.4\% for \qfgz=3 (Figure~\ref{fig:reliability}).
A visualization of all grism spectra with \qfgz$\geq3$ is shown in Figure~\ref{fig:grismplot}.

Given the large number of reliable redshifts in this survey and lack of target selection bias for grism data, \survey\ provides an excellent test of photometric redshift codes.
We compare these high quality redshifts to the COSMOS2020 CLASSIC \zp\ from both LePhare (lp\_zBEST) and EAZY (ez\_z500), considering those objects which have $0.15<$\zp$<8$ in each case.
Both codes do an excellent job on average as shown in Figure~\ref{fig:zpzg}.
While EAZY has very slightly smaller values for median offset and scatter, it also has a slightly larger outlier fraction.

Finally, since we have reprocessed and analyzed data from 3D-\HST\ in concert with the new \survey\ observations, we compare the redshifts for those objects in the 3D-\HST\ footprint as determined by the two survey groups.
We begin by taking objects with reliable redshifts (\qfgz~=3,4,5 for \survey\ and $use=1$ for 3D-\HST) and comparing to secure spectroscopic redshifts (\qfs~=3,4).
Both surveys do quite well and while the 3D-\HST\ team redshifts have a slightly smaller median offset ($\Delta z / (1+z)=-0.0027$ vs. 0.0047), the \survey\ team redshifts have a smaller scatter ($\sigma_{\rm NMAD}=0.0159$ vs. 0.0213; Figure~\ref{fig:zgzs}).
Next we perform a direct comparison between objects with usable grism redshifts in both surveys and, as expected given the previous test, find good agreement between the two surveys (Figure~\ref{fig:zgcomp}).
The scatter between the two sets of grism redshifts decreases significantly when objects with \zg~$<0.7$ are excluded, as these galaxies typically have featureless continua in G141.
We thus conclude that our redshift processing and analysis methodology is sound.

\begin{figure*}
    \includegraphics[width=\textwidth, trim=0in 0in 0in 0in]{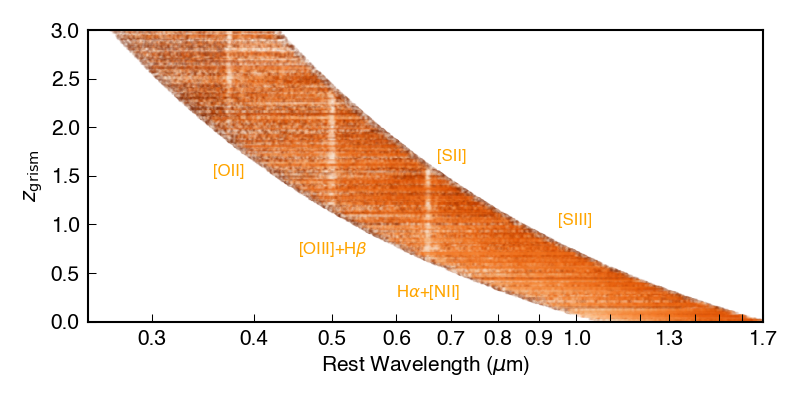}
    \caption{A visualization of grism spectra for objects with \qfgz~$\geq3$. Strong emission lines such as \OII, \OIII+\Hbeta, and \Halpha+[NII] are clearly visible.}
    \label{fig:grismplot}
\end{figure*}

\begin{figure*}
    \includegraphics[width=\textwidth, trim=0in 0in 0in 0in]{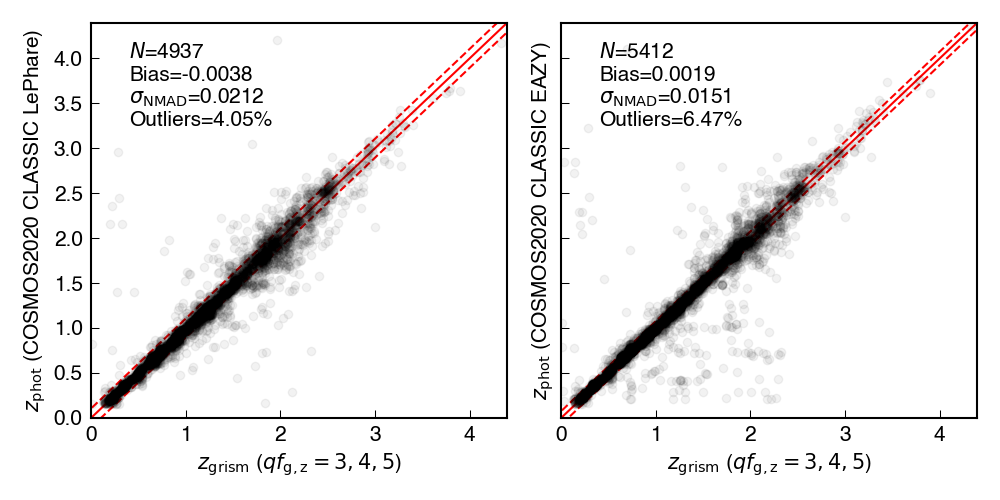}
    \caption{Comparison of photometric redshifts to high-quality grism redshifts for sources with \zp~$>0.15$. Photometric redshifts are from the COSMOS2020 CLASSIC catalog, specifically the `lp\_zBEST' (left) and `ez\_z500' (right) entries. Both codes perform well, with EAZY resulting in  smaller bias, smaller scatter, and a higher fraction of outliers - those greater than $5\sigma_{\rm NMAD}$ from agreement. These limits are shown by the dashed red lines, while one-to-one agreement is the solid red line.}
    \label{fig:zpzg}
\end{figure*}

\begin{figure*}
	\includegraphics[width=\textwidth, trim=0in 0in 0in 0in]{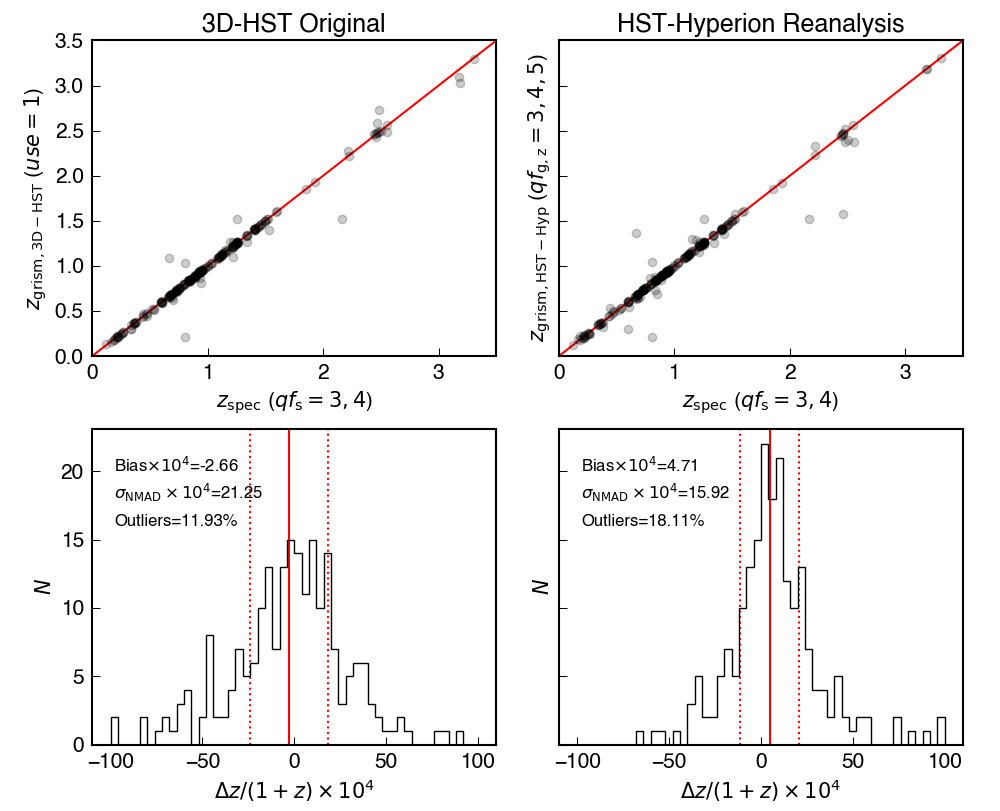}
    \caption{Comparison of high-quality grism redshifts to high-quality ground-based spectroscopic redshifts derived for 243 objects in the 3D-\HST\ footprint. The original redshifts from the 3D-\HST\ survey are on the left, while those from the reanalysis by the \survey\ survey are on the right. The distribution of $\Delta z/(1+z)$ is shown on the bottom row.}
    \label{fig:zgzs}
\end{figure*}

\begin{figure*}
	\includegraphics[width=\textwidth, trim=0in 0in 0in 0in]{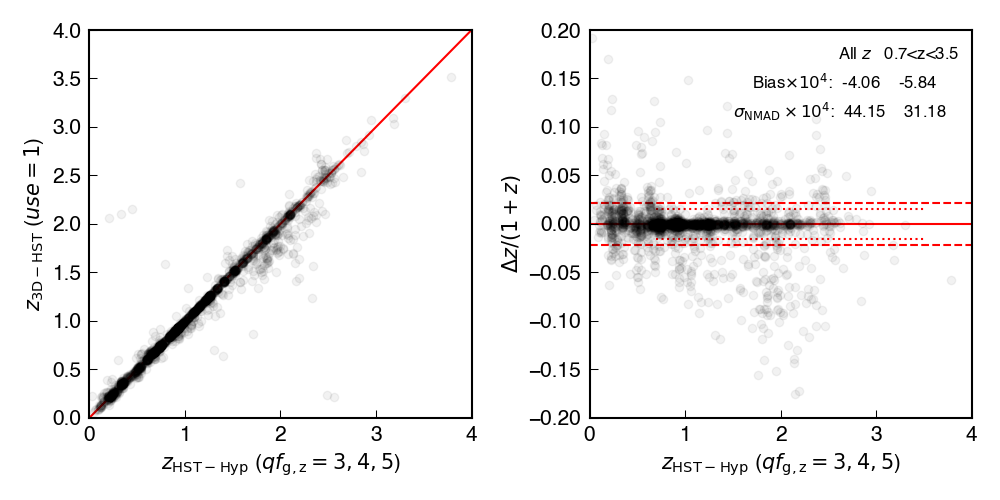}
    \caption{Comparison of grism redshifts derived for objects in the 3D-\HST\ footprint by the \survey\ and 3D-\HST\ surveys. The solid red line is the 1:1 relation, dashed lines represent $\pm5\sigma_{\rm NMAD}$ derived from targets at all redshifts, and dotted lines represent $\pm5\sigma_{\rm NMAD}$ derived from targets at $0.7<z<3.5$.}
    \label{fig:zgcomp}
\end{figure*}

\section{Emission Lines}\label{Sec:EL}

\subsection{Modeling Emission Lines with \grizli}

As mentioned in Section~\ref{Sec:DRgrizli}, \grizli\ uses combined fitting to photometric and grism data to obtain a redshift using spectral models which consider several (often) strong emission lines in fixed ratios.
Subsequently, a best-fit model is obtained by fixing the redshift and allowing additional lines to be fit without the same line ratio constraints imposed.
While this allows for less common line ratios to be fit when observed in the data, noisy data can also lead to physically unlikely or impossible line ratios.
However, fitting the lines and stellar continuum simultaneously is important for modeling emission lines that lie on top of stellar absorption features, \eg\ the Balmer series lines.
Here we only consider the strongest rest-frame optical lines, \OII, \Hbeta, \OIII, and \Halpha, which in many cases have sufficient signal to avoid significant flux errors and for which stellar absorption is generally minimal or non-existent.
The distribution of line fluxes and rest-frame equivalent widths for these lines are shown in Figure~\ref{fig:lines}.
The best-fit model continuum is used in calculating these values.

We estimate the line sensitivity of the \survey\ survey by plotting the signal-to-noise ratio (SNR) of the strong emission lines as a function of line flux.
All grism sources with \qfgz~$\geq 3$ and a redshift in which a strong line will fall in the grism (using $1.10<\lambda/\mu m<1.65$) are considered.
We then find the line flux at which the median of SNR is unity (Figure~\ref{fig:EL_SNR}), which corresponds to a line flux of \mbox{$8.8\times 10^{-18}$ erg s$^{-1}$ cm$^{-2}$}.

While this is in excellent agreement with the 3D-\HST\ survey, which found a limit of \mbox{$8\times 10^{-18}$ erg s$^{-1}$ cm$^{-2}$} with identical grism depth, the 3D-\HST\ value is for a source with radius of 5 pixels and a line at 1.5 $\mu m$.
The sensitivity of the grism is wavelength dependent, and the line flux uncertainty depends upon the extraction aperture size.
The throughput sensitivity factor is dominant, as shown in \citet{Momcheva2016}, and we test the dependance of our line sensitivity calculation on this factor.
The WFC3/G141 sensitivity is relatively flat and maximal across the range $1.30<\lambda/\mu m<1.55$, corresponding to redshifts of \mbox{$2.49<z_{\rm [OII]}<3.16$}, \mbox{$1.67<z_{\rm H\beta}<2.19$}, \mbox{$1.60<z_{\rm [OIII]}<2.10$}, and \mbox{$0.98<z_{\rm H\alpha}<1.36$} for the strong rest-frame emission lines.
Considering grism sources with \qfgz~$\geq 3$ in these redshift ranges only yields a sensitivity of \mbox{$7.1\times 10^{-18}$ erg s$^{-1}$ cm$^{-2}$}, which is a small improvement in sensitivity as expected.

\subsection{Emission Line Catalog}

In addition to the redshift catalog described above, we also release an emission line catalog.
This catalog is line-matched to the redshift catalog, and includes the same grism object ID, coordinates, \hstmag, grism redshift, and grism redshift quality flag for ease of use.
The equivalent widths and integrated line fluxes for \OII, \Hbeta, \OIII, and \Halpha\ are also included, as listed in Table~\ref{tab:ELcat}.

During visual inspection of the model fitting from \grizli, the ability of the model to reproduce the observed data is a consideration.
Although in this way the line fluxes and equivalent widths have been visually checked, no further inspection or tests of these values as provided by \grizli\ have been performed outside of the analysis in this work.

\subsection{Comparison to 3D-\HST\ Results}

Similar to the comparison of redshifts between sources included in both \survey\ and 3D-\HST\ performed in Section~\ref{Sec:zg_comp}, we also compare the emission line properties for common sources in the two surveys.
As in the redshift comparison we only include \survey\ sources with \qfgz~$\geq3$ and 3D-\HST\ sources with $use=1$, with object matching performed by selecting the nearest match with a projected offset of $<0.5$\arcsec.
We do not enforce grism redshift equality across the two surveys, but as previously shown, these differences are typically miniscule and will not lead to misidentification of strong emission lines.

In Figure~\ref{fig:lfcomp} we show the comparison of line fluxes for the strong optical emission lines.
3D-\HST\ tends to measure slightly smaller line fluxes than \survey: $\sim0.05$ dex for \OII\ and \OIII, $\sim0.12$ dex for \Halpha.
This is potentially due to differences in the photometry used for fitting in conjunction with the grism data, which results in a different stellar continuum fit-  3D-\HST\ used CANDELS and 3D-\HST\ imaging, whereas \survey\ uses the COSMOS2020 catalogs.
Nonetheless, the offsets are generally small and do not show a dependence on line flux.
Additionally, for \OII\ and \OIII\ the median offset is significantly smaller than the scatter of the distribution, while for \Halpha\ they are more comparable.
\Hbeta\ is the weakest line on average and, as such, is more dependent on the stellar continuum absorption strength than \Halpha, resulting in larger scatter.

Somewhat similar results are found when looking at equivalent widths (Figure~\ref{fig:lewcomp}), with a scatter in the relation larger than the median offset.
3D-\HST\ however has larger equivalent widths than \survey\ for \OII, \Hbeta, and \OIII.
Combined with the lower line fluxes, this is suggestive of a normalization difference between the two surveys, either due to the different photometry used, different models incorporated, or perhaps due to processing differences from different versions of \grizli.

\begin{table}
	\centering
	\caption{Columns in the \survey\ emission line catalog, where the lines considered (XX) are \OII, \Hbeta, \OIII, and \Halpha.}
	\label{tab:ELcat}
	\begin{tabular}{ll}
		\hline
		Column	& Description \\
		\hline
            ID\_g	&   Target ID in \survey \\
            RA	&   Right Ascension of the grism source (J2000)\\
            DEC  	&   Declination of the grism source (J2000)\\
            mag	&   \hstmag\ of the grism source\\
            z\_g   	&   Grism redshift from \survey \\
            qf\_gz	&   Grism redshift quality flag \\
            EW50\_XX	& Median of equivalent width pdf for XX line\\
            EW16\_XX	& 16$^{\rm th}$ percentile of EW pdf for XX line\\
            EW84\_XX	& 84$^{\rm th}$ percentile of EW pdf for XX line\\
            F\_XX	& Modeled integrated line flux for XX line\\
            eF\_XX	& Uncertainty on line flux for XX line\\
	\end{tabular}
\end{table}

\begin{figure*}
    \includegraphics[width=\textwidth, trim=0in 0in 0in 0in]{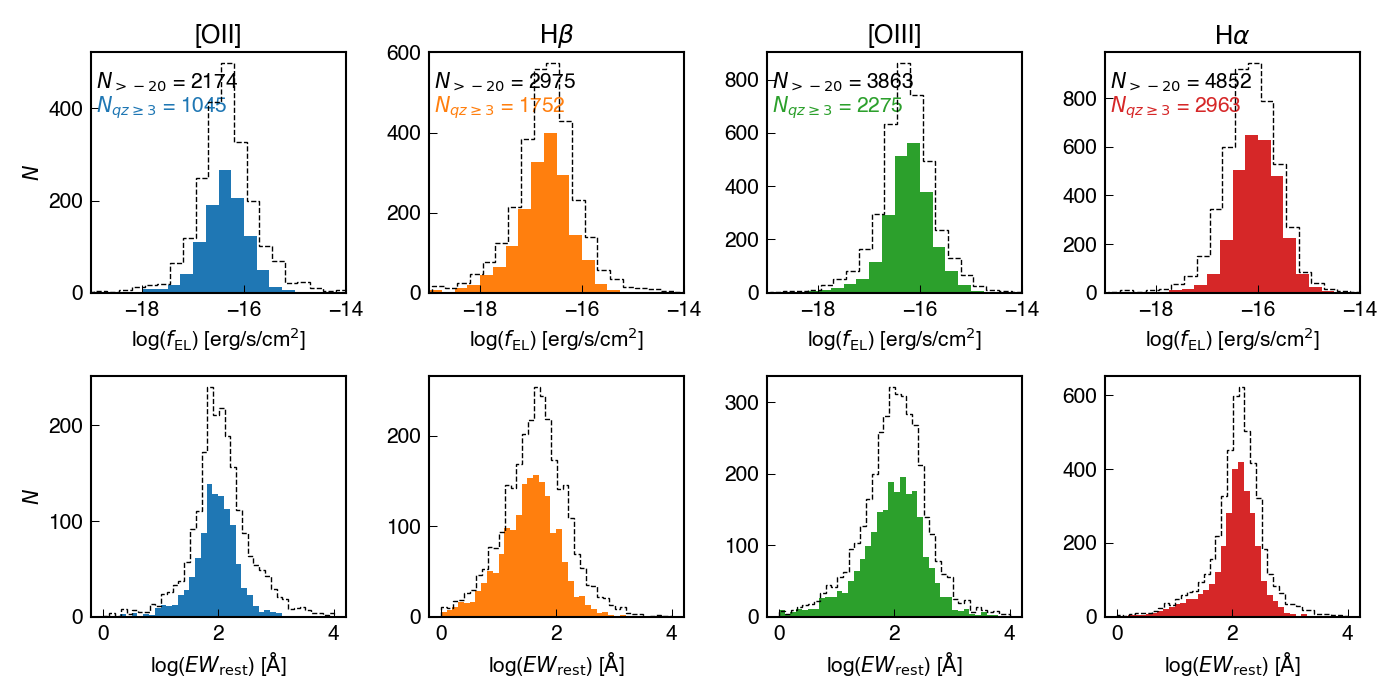}
    \caption{Histograms of the strong emission line fluxes (top) and rest-frame equivalent widths (bottom) from \survey. From left to right these include \OII, \Hbeta, \OIII, and \Halpha. Black dashed histograms represent results from all sources with the line falling into the grism wavelength range at the assigned redshift, while the colored histograms are limited to such objects with \qfgz~=~3,4,5.}
    \label{fig:lines}
\end{figure*}

\begin{figure}
    \includegraphics[width=0.45\textwidth, trim=0in 0in 0in 0in]{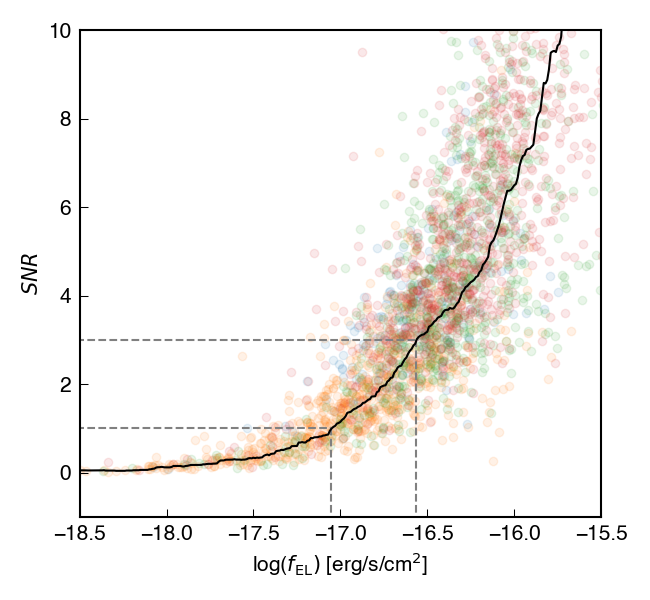}
    \caption{The signal-to-noise ratio of the strong emission line fluxes from \survey\ as a function of line flux. Only sources with \qfgz~=~3,4,5 and a measured emission line flux in the WFC3/G141 wavelength range for \OII\ (blue), \Hbeta\ (green), \OIII\ (orange), and \Halpha\ (red) are included. The black line shows the 50th percentile SNR in a narrow line flux bin. This corresponds to $8.8\times 10^{-18}$ erg s$^{-1}$ cm$^{-2}$ at $1\sigma$.}
    \label{fig:EL_SNR}
\end{figure}

\begin{figure}
    \includegraphics[width=0.5\textwidth, trim=0in 0in 0in 0in]{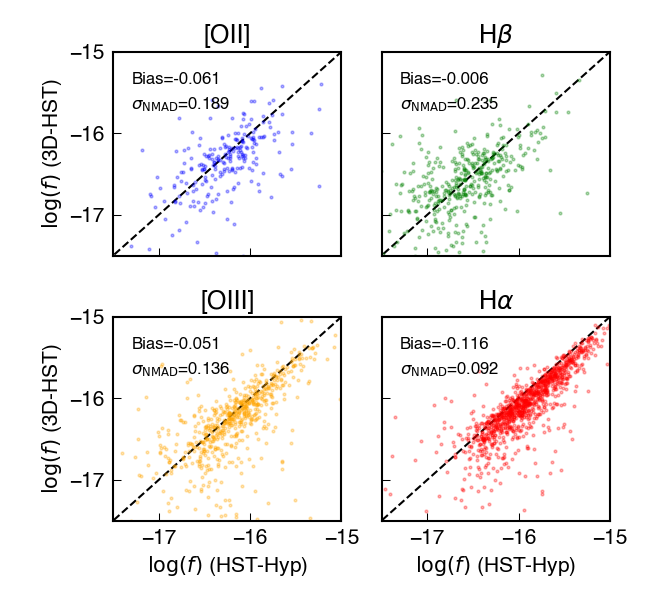}
    \caption{Comparison of line fluxes for objects in the 3D-$HST$ footprint. Original values from 3D-$HST$ are on the y-axes and reanalyzed values from \survey\ are on the x-axes. Clockwise from top left these include \OII, \Hbeta, \OIII, and \Halpha.}
    \label{fig:lfcomp}
\end{figure}

\begin{figure}
    \includegraphics[width=0.5\textwidth, trim=0in 0in 0in 0in]{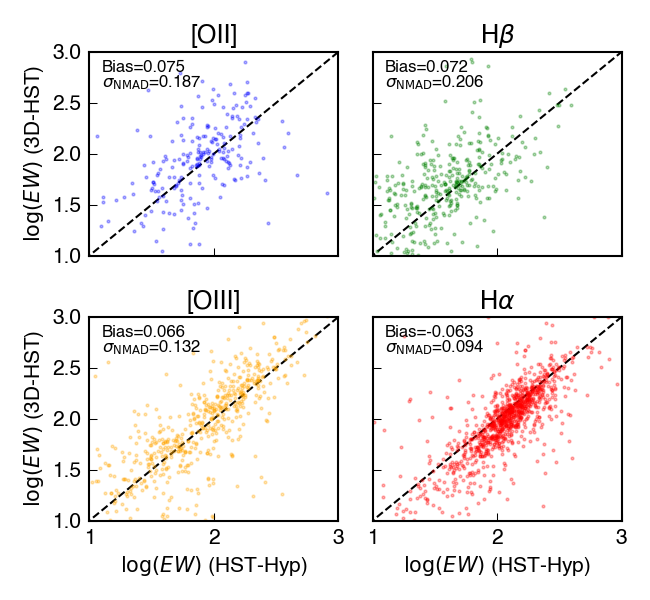}
    \caption{Comparison of line equivalent widths for objects in the 3D-$HST$ footprint. Original values from 3D-$HST$ are on the y-axes and reanalyzed values from \survey\ are on the x-axes. Clockwise from top left these include \OII, \Hbeta, \OIII, and \Halpha.}
    \label{fig:lewcomp}
\end{figure}

\section{Hyperion: An Updated picture}\label{Sec:Hyp}

The Hyperion proto-supercluster at $z\sim2.45$ has been spectroscopically observed with ground-based field surveys including zCOSMOS, DEIMOS10k, and VUDS, as well as targeted observations published in \citet{Casey2015, Diener2015, Chiang2015, Wang2016}.
The C3VO team has also performed targeted spectroscopic follow-up of Hyperion with Keck/MOSFIRE before also analyzing the \HST/WFC3/G141 data discussed herein.
Combined, these spectroscopic efforts have resulted in over 650 galaxies with high-quality spectroscopic redshifts in the cube spanning \mbox{$149.60<$~R.A.~$<150.60$}, \mbox{$1.75<$~Dec.~$<2.75$}, and \mbox{$2.35<z<2.65$}.
This includes 196 galaxies with high-quality grism redshifts from \survey, of which 149 are new (Figure~\ref{fig:members}).

Using the three-dimensional overdensity maps from \citep{Cucciati2018} and several spectroscopic catalogs, we confirm new member galaxies in most of the $\geq 5\sigma$ overdense peak regions, which range from $2.40<z<2.53$ (Figure~\ref{fig:peak_members}; Table~\ref{tab:pm}).
However, we note that the galaxies with confirmed redshifts in $2.35<z<2.65$ may suggest additional structure.
A reconstruction of the density map of the field may be beneficial, and we will present such a map in a future work.
We also note that an analysis of a large spectroscopic catalog in COSMOS suggests the presence of 3 overdensity peaks in the system \citep{Khostovan2025}.

Even with the reduction of spectroscopic targeting bias provided by the grism observations, the densest region of the system appears to be located at (R.A., Dec., $z$) = (150.235, 2.335, 2.506).
This corresponds to the structure identified in \citet{Wang2016} and Peak [5] of \citet{Cucciati2018}.
The region is highly compact, with an X-ray detection and a velocity dispersion of $\sigma_v=503$ km/s as determined in both works.
While the X-ray detection is both secure and extended, whether it originates from hot gas in the intracluster medium \citep{Wang2016} or from inverse Compton scattering from a radio-loud AGN which has since gone quiet \citep{Champagne2021} is unclear.

\begin{table}
	\centering
	\caption{The number of confirmed sources in each region of peak overdensity from \citet{Cucciati2018}.}
	\label{tab:pm}
	\begin{tabular}{lcccr} 
		\hline
		  C+18 Peak	& R.A. 	& Dec. & z  & $N_{spec-z}$ \\
		\hline
            1 (Theia)   & 150.0937 &  2.4049 &  2.468  & 32\\
            2 (Eos)     & 149.9765 &  2.1124 &  2.426  & 8\\
            3 (Helios)  & 149.9996 &  2.2537 &  2.444  & 23\\
            4 (Selene)  & 150.2556 &  2.3423 &  2.469  & 16\\
            5           & 150.2293 &  2.3381 &  2.507  & 14\\
            6           & 150.3316 &  2.2427 &  2.492  & 2\\
            7           & 149.9581 &  2.2187 &  2.423  & 3\\
		\hline
	\end{tabular}
\end{table}

\begin{figure*}
    \includegraphics[width=\textwidth, trim=0in 0in 2in 0in]{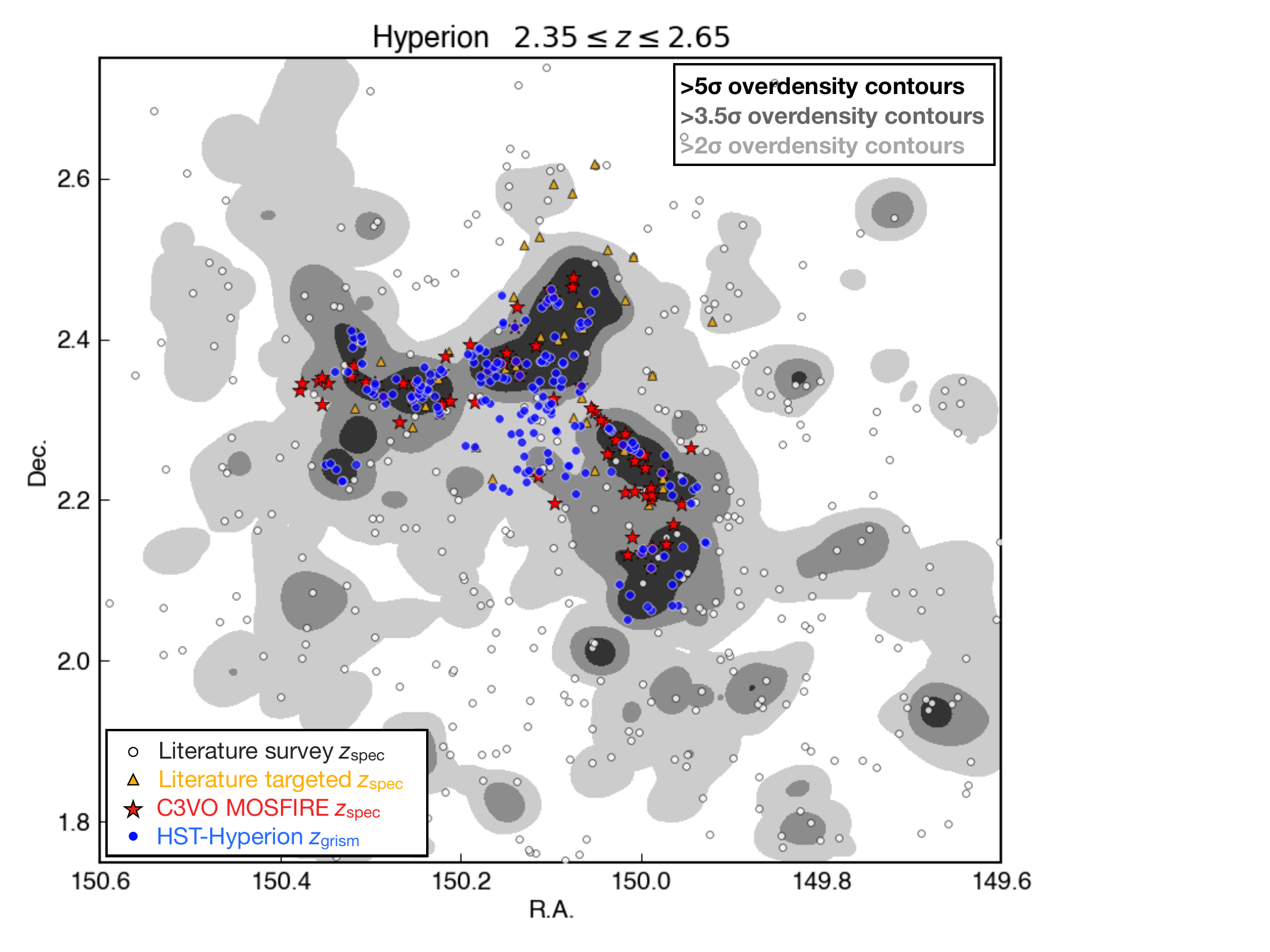}
    \caption{Spectroscopically confirmed sources in the region of the Hyperion proto-supercluster. The projected $5\sigma$, $3.5\sigma$, and $2\sigma$ overdensity contours from \citet{Cucciati2018} are shown in shades of gray. Spectroscopic redshifts from the zCOSMOS, DEIMOS10k, and VUDS field surveys with $2.35<z_{\rm spec}<2.65$ and \qfs~=~3,4 are shown as black rings, while spectroscopically confirmed members from targeted works are shown as gold triangles \citep{Casey2015, Diener2015, Chiang2015, Wang2016}.
    Members identified in the C3VO MOSFIRE followup of Hyperion with the same cuts are shown as red stars. \survey\ sources with $2.35<z_{\rm grism}<2.65$ and \qfgz~=~3,4,5 are blue. The \survey\ program has resulted in 192 potential members in this redshift range with high-quality grism redshifts.}
    \label{fig:members}
\end{figure*}

\begin{figure*}
    \includegraphics[width=\textwidth, trim=0in 0in 2in 0in]{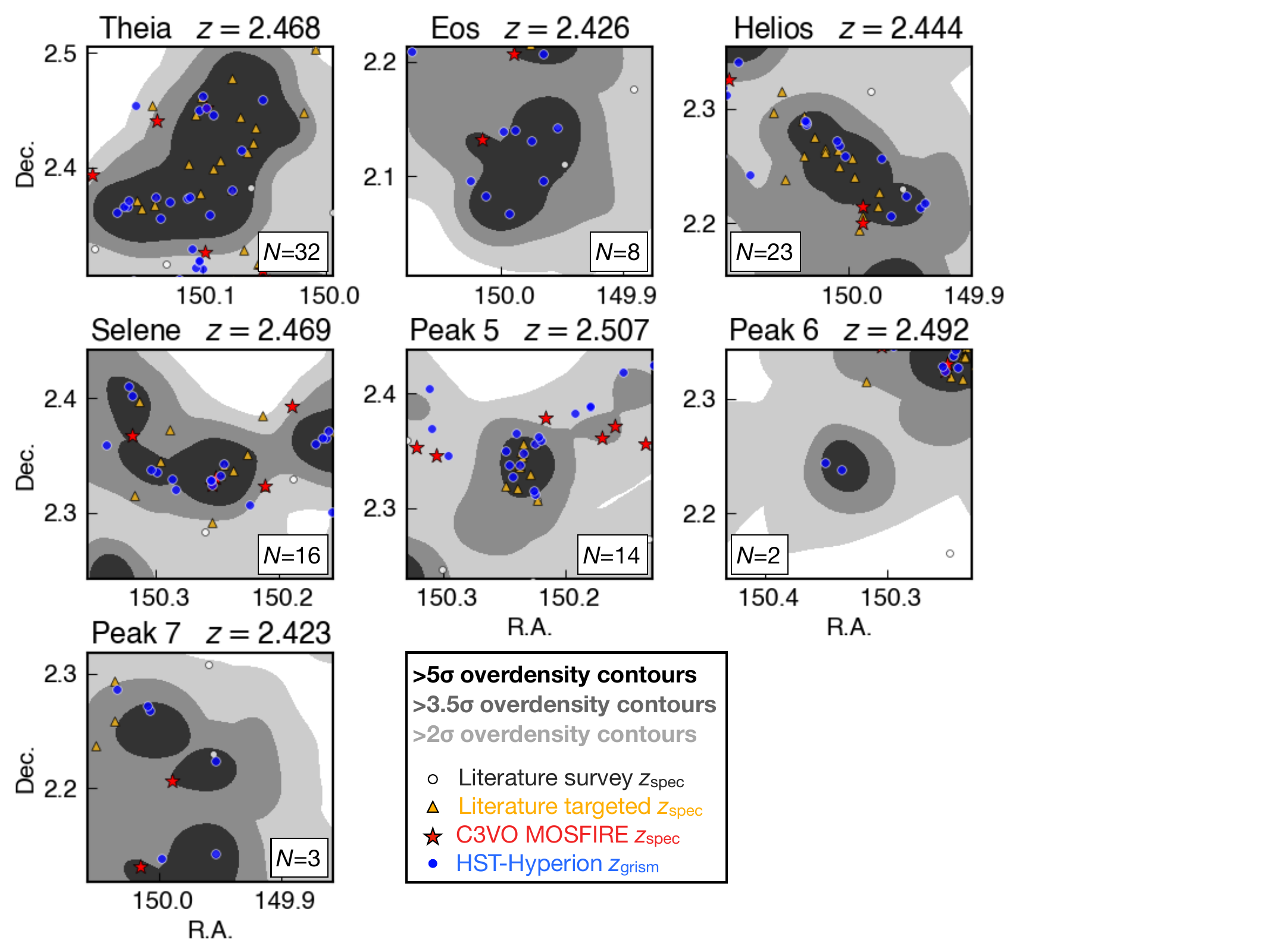}
    \caption{The seven $>5\sigma$ overdensity peaks identifed in \citet{Cucciati2018}. The contours and marker shapes/colors are the same as in Figure~\ref{fig:members}. Only those galaxies within the redshift range of the peak region ($z_{\rm max} - z_{\rm min}\sim 0.03-0.05$) are included.}
    \label{fig:peak_members}
\end{figure*}

\section{Summary}

In this work, we have described the observations, data reduction process, and characteristics of the 50-orbit \survey\ WFC3/G141 survey.
Combining these grism observations with those from the 3D-\HST\ survey and the extensive photometric observations in the COSMOS2020 catalog, we have visually inspected $>12000$ sources with \hstmag~$\leq 25$ and recovered reliable redshifts for 5629
sources and emission line fluxes with a $1\sigma$ sensitivity of $8.8\times10^{-18}$ erg s$^{-1}$ cm$^{-2}$.

We compared our results to those from the 3D-\HST\ survey for overlapping sources and found good agreement between not only redshifts but also emission line models.
Comparison to high-quality ground-based spectroscopic redshifts also shows excellent agreement, with median offset of $\Delta z = 0.0007$ and $\sigma_{\rm NMAD}\sim0.0011$ for sources with $z>0.7$, where strong emission lines fall in the grism wavelength window.
We also use the \survey\ grism redshifts as a check of photometric redshift reliability.
Redshifts from the COSMOS2020 CLASSIC catalog have $\sigma_{\rm NMAD}\sim0.021$ (LePhare) and $\sigma_{\rm NMAD}\sim0.015$ (EAZY) with no significant biases.

The grism observations result in 196 spectroscopic redshifts over $2.35<z<2.65$, of which 149 are unique from other spectral catalogs we consider.
These include 125 spectroscopically-confirmed members of the Hyperion system at $2.40<z<2.53$ and an additional 71 confirmed galaxies within $2.35<z<2.65$ in regions which may be associated with the Hyperion system.
Combined with other spectroscopic observations, the total number of galaxies with confirmed redshifts in the range $2.35<z<2.65$ over the 1~deg$^2$ region centered on Hyperion and with 
such members is over 650 galaxies.

The Hyperion system, which consists of at least one potentially virialized substructure and multiple other overdense peaks, continues to be one of the premier systems in which to study the formation and collapse of structure and its effects on galaxy evolution.
Future observations with facilities such as ALMA and {\it JWST} (\eg\ COSMOS-3D, PI: Kakiichi, PID 5893) will provide synergistic data with which to explore these questions.
To that end, we release a redshift catalog, a stellar population catalog}, and an emission line catalog with \survey\ results for public use.

\clearpage

\begin{acknowledgments}
We dedicate this work to the memory of Bianca Garilli, who played an integral role in the VUDS and zCOSMOS surveys utilized in this work, and was an excellent scientist and person.
This research is based on observations made with the NASA/ESA Hubble Space Telescope obtained from the Space Telescope Science Institute, which is operated by the Association of Universities for Research in Astronomy, Inc., under NASA contract NAS 5-26555. These observations are associated with program GO-16684.
The authors wish to recognize and acknowledge the very significant cultural role and reverence that the summit of Maunakea has always had within the indigenous Hawaiian community. 
We are most fortunate to have the opportunity to conduct observations from this mountain.
This work is also based on observations collected at the European Southern Observatory under ESO programmes 175.A-0839, 179.A-2005, and 185.A-0791, as well as work supported by the National Science Foundation under Grant No. 1908422.
BF also acknowledges support from JWST-GO-02913.001-A and the hospitality of Rutgers University.
The authors thank the anonymous referee for their feedback and contributions which have improved the manuscript.

\end{acknowledgments}

\facilities{\HST(WFC3), Keck(MOSFIRE), Keck(DEIMOS), VLT(VIMOS)}

\software{astropy \citep{Astropy2013, Astropy2022},
\grizli\ \citep{Brammer2021}}

\bibliography{library}{}
\bibliographystyle{aasjournal}

\appendix
\renewcommand\thefigure{\thesection.\arabic{figure}}    

\section{Data Reduction and Classification Verification Using Half-Depth Observations}
\setcounter{figure}{0}

One of the pointings (WFC3-PEAK4-2) had several observation attempts produce unusable data due to fine guidance sensor failures causing pointing drifts during exposure.
A successful observation was not acquired until December 2024 and as such, analysis initially proceeded using integrations of effectively half depth for this pointing.
Upon successful completion of observations, this region was re-reduced and sources re-classified for inclusion in the final catalogs.
This enables us to quantify the effect of doubling the depth of observations, as well as compare classifications between additional classifiers.

\begin{figure*}
    \includegraphics[width=0.5\textwidth, trim=0in 0in 0in 0in]{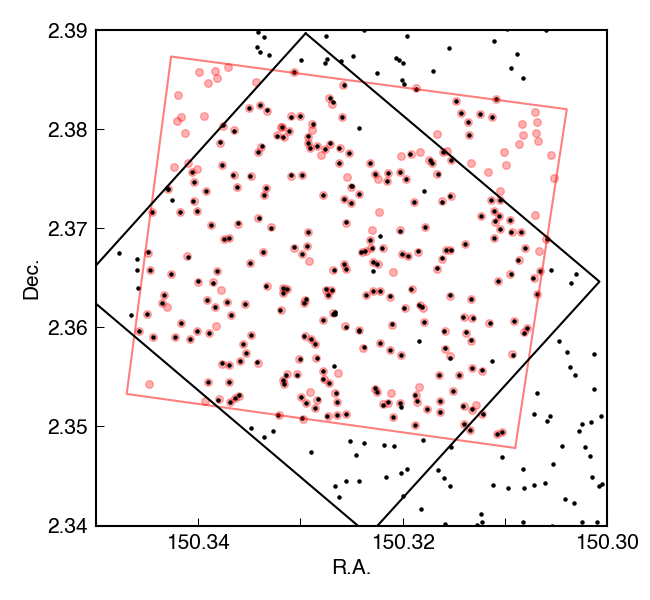}
    \caption{A comparison of the position angles used to obtain data for WFC3-PEAK4-2. The field ov view for the initial data is shown as a black rectangle, while the pointing for the final observations is shown as a red rectangle. Sources detected in the initial classification are shown in black (half depth within the black rectangle), while sources detected within the red rectangle (considering all data) are shown in red.}
    \label{fig:newobs_sources}
\end{figure*}

\begin{figure*}
    \includegraphics[width=\textwidth, trim=0in 0in 0in 0in]{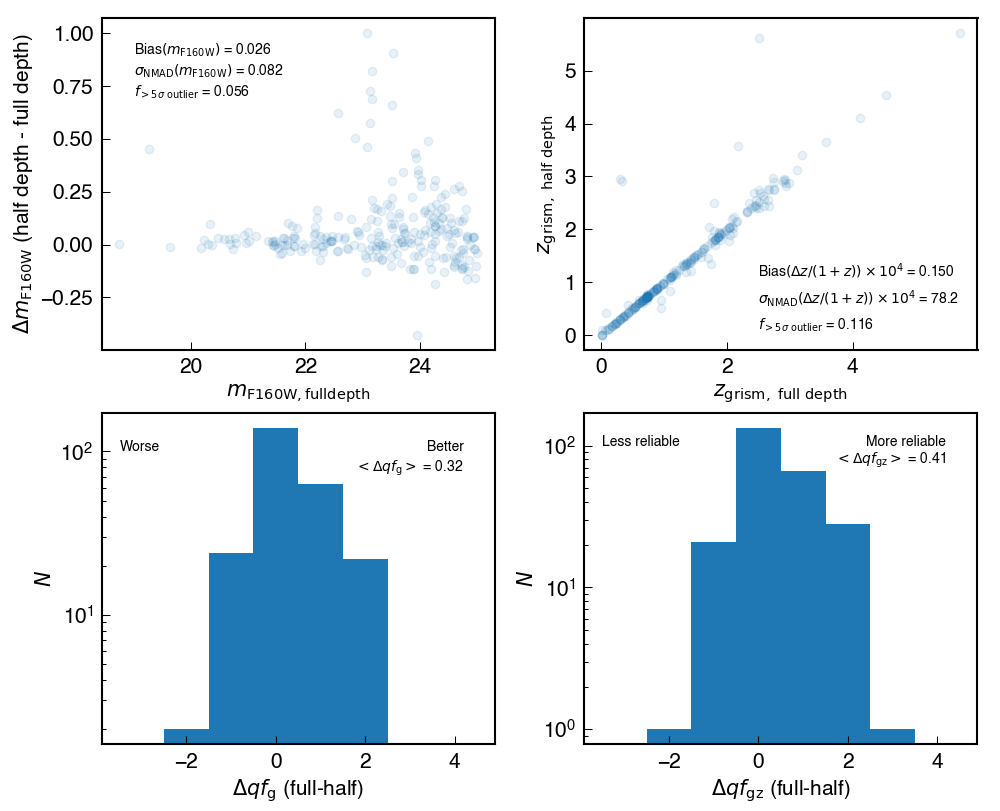}
    \caption{A comparison in derived values for sources in WFC3-PEAK4-2 when considering data at half-depth and full-depth. Clockwise from top left: Difference in $m_{\rm F160W}$, difference in modeled grism redshift, difference in grism redshift quality flag, and difference in grism data quality flag.}
    \label{fig:newobs_stats}
\end{figure*}

Figure~\ref{fig:newobs_sources} shows the change in position angle required to obtain the second half of the integration for WFC3-PEAK4-2.
Reanalysis of the data at full depth produced 302 sources in the footprint of the new data, with 250 in common with the analysis of the previous data.
There are 24 new sources detected in the overlapping region (detected at full depth, but not half depth) and 8 sources which were detected at half depth but not full depth (likely spurious noise detections).
Figure~\ref{fig:newobs_stats} shows that agreement between magnitudes and redshifts determined from data at different depths is generally quite good, though the level of confidence in model redshift fits increases with the additional depth.

\section{Identification of Overdensities}
\setcounter{figure}{0}

\begin{figure*}
    \includegraphics[width=\textwidth, trim=0in 0in 0in 0in]{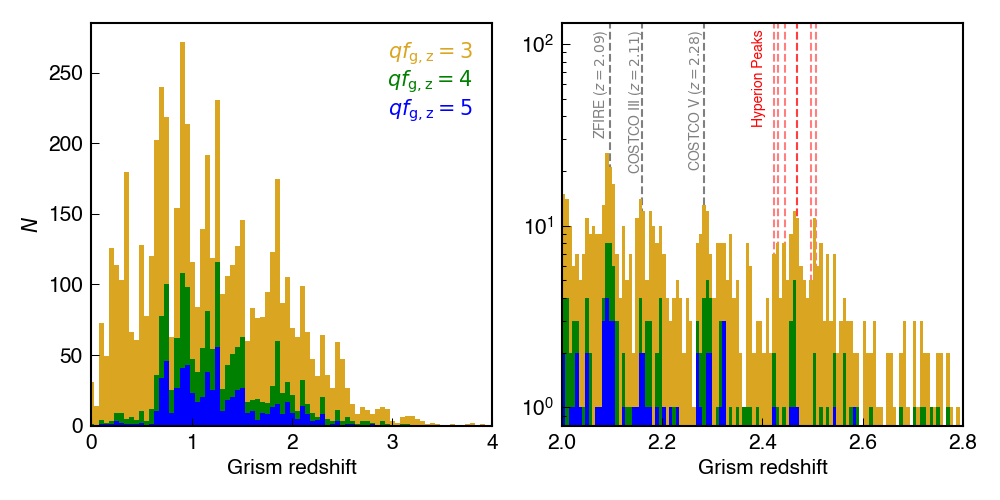}
    \caption{Redshift histogram of the high-quality grism redshifts analyzed in this work. \textbf{Left:} All sources colored by \qfgz. \textbf{Right:} All sources in $2.0<z<2.8$, with the redshift of literature overdensities shown as vertical dashed lines.}
    \label{fig:zhist}
\end{figure*}

The identification of candidate overdensities in a field is often aided by the identification of peaks in redshift histograms.
While we caution that the clustering of sources at a similar redshift must be considered, and that this method is unlikely to yield a pure or complete sample of structures, it is worthwhile to see whether such a wide field grism survey can recover such structures.
The left panel of Figure~\ref{fig:zhist} shows the histogram of all sources with a grism redshift and \qfgz~$\geq 3$. The right panel shows a zoom-in of the region $2<z<2.8$.
In addition to Hyperion, three other overdensities referenced in \citet{Hung2025} within $2<z<2.8$ lie in the \survey\ pointings.
These include the ZFIRE protocluster at $z=2.09$ \citep{Yuan2014} and two protoclusters from \citet{Ata2022} - COSTCO III ($z=2.11$) and COSTCO V ($z=2.28)$.
The redshifts of these three overdensities and the Hyperion peaks are all associated with increases in the number of high-quality grism redshifts.

\end{document}